\documentclass[9pt,twoside]{pnas-new}
\templatetype{pnasresearcharticle} % Choose template 
\usepackage{multicol}
\usepackage{flushend}
\usepackage{geometry}

\usepackage{float}

\title{Bacteria hinder large-scale transport and enhance small-scale mixing in time-periodic flows}

\author[a]{Ranjiangshang Ran}
\author[a]{Quentin Brosseau} 
\author[a]{Brendan C. Blackwell} 
\author[a]{Boyang Qin}
\author[a]{Rebecca Winter} 
\author[a,1]{Paulo E. Arratia}

\affil[a]{Department of Mechanical Engineering and Applied Mechanics, University of Pennsylvania, Philadelphia, PA 19104}

\leadauthor{Ran}

\begin{abstract}
Understanding mixing and transport of passive scalars in active fluids is important to many natural (e.g. algal blooms) and industrial (e.g. biofuel, vaccine production) processes. Here, we study the mixing of a passive scalar (dye) in dilute suspensions of swimming \textit{Escherichia coli} in experiments using a two-dimensional (2D) time-periodic flow and in a simple simulation. Results show that the presence of bacteria hinders large scale transport and reduce overall mixing rate. Stretching fields, calculated from experimentally measured velocity fields, show that bacterial activity attenuates fluid stretching and lowers flow chaoticity. Simulations suggest that this attenuation may be attributed to a transient accumulation of bacteria along regions of high stretching. Spatial power spectra and correlation functions of dye concentration fields show that the transport of scalar variance across scales is also hindered by bacterial activity, resulting in an increase in average size and lifetime of structures. On the other hand, at small scales, activity seems to enhance local mixing. One piece of evidence is that the probability distribution of the spatial concentration gradients is nearly symmetric with a vanishing skewness. Overall, our results show that the coupling between activity and flow can lead to nontrivial effects on mixing and transport. 

\end{abstract}

\dates{This manuscript was submitted to \textit{PNAS} on July 30, 2021.}
\doi{${}^{\text{1 }}$To whom correspondence should be addressed. E-mail: parratia@seas.upenn.edu}

\begin{document}

\maketitle
\thispagestyle{firststyle}
\ifthenelse{\boolean{shortarticle}}{\ifthenelse{\boolean{singlecolumn}}{\abscontentformatted}{\abscontent}}{}

% If your first paragraph (i.e. with the \dropcap) contains a list environment (quote, quotation, theorem, definition, enumerate, itemize...), the line after the list may have some extra indentation. If this is the case, add \parshape=0 to the end of the list environment.

\begin{multicols}{2}
Swimming microorganisms often live in environments with fluid flows across a range of length scales, from natural habitats like oceans to industrial biofuel plants to human intestines \cite{Vogel_book_1983}. Flow can exert forces and torques on microorganisms, which can affect their locomotion and transport \cite{Stocker_Annual_Review_2019}. Microorganisms, conversely, adapt their swimming directions and take advantage of flow gradients in order to forage and reproduce \cite{Taylor_Science_2012}. This coupling between activity (i.e. swimming motion) and flow leads to intriguing physical phenomena that are not seen for passive particles, such as rheotaxis \cite{Hill_PRL_2007, Marcos_PNAS_2012}, gyrotaxis \cite{DeLillo_PRL_2014, Borgnino_JFM_2018}, and flow-induced chemotaxis \cite{Locsei_BMB_2009, Shapiro_PNAS_2014}.

Even simple flows, such as shear flows, can profoundly alter the movement of microswimmers \cite{rusconi_NatPhys_2014,Stocker_JRSI_2015, Durham_Science_2009}. Experiments show motile bacteria can drift across streamlines out of the plane of shear \cite{Marcos_PNAS_2012}, escape the low-shear regions and get trapped in high-shear regions \cite{rusconi_NatPhys_2014}. Motile phytoplankton are found to deplete or accumulate in regions of different shear rates \cite{Stocker_JRSI_2015}, and form intense cell assemblages called ``thin layers''\cite{Durham_Science_2009}. Under quiescent flow condition, microswimmers can still induce fluctuating velocity fields in the fluid by moving collectively \cite{Guasto_PRL_2010, Rushkin_PRL_2010}. These self-induced flows can lead to unique properties in fluids laden with active particles, or namely ``active suspensions''. Examples include enhanced Brownian diffusivity \cite{Wu_PRL_2000, Kurtuldu_PNAS_2011, Poon_PRE_2013}, active fluid transport and mixing \cite{Short_PNAS_2006,Saintillan_PRL_2008, Pushkin_PRL_2013}, and even the possibility of work extraction \cite{Sokolov_PNAS_2010, DiLeonardo_PNAS_2010}. 

In more complex flows, the behavior of microswimmers shows rich dynamics but is much less understood. In turbulent flows, gyrotactic swimmers are found to cluster in small-scale patches \cite{Durham_NC_2013, DeLillo_PRL_2014}, and gather in regions of positive velocity gradients \cite{Gustavsson_PRL_2016}. Numerical simulations in 3D isotropic turbulence show that elongated swimmers preferentially align with flow velocity \cite{Borgnino_PRL_2019}, and clustering and patchiness are greatly reduced \cite{brandt_JFM2014}. In chaotic flows, simulations show that rod-like swimmers can be trapped or expelled by elliptic islands \cite{Torney_PRL_2007}, i.e. KAM tori \cite{Ottino1989}, depending on their shapes and swimming speeds. The trapping of particles in elliptic islands can even lead to a reduction in swimmer transport \cite{Khurana_PRL_2011, Khurana2012}. Recently, a study on the steady flow in model porous media show that bacteria align and accumulate with Lagrangian structures \cite{Guasto_PNAS2019}. However, how these dynamics affect mixing and transport in flows exhibiting chaotic advection are yet to be experimentally tested. Moreover, most simulations treat swimmers as self-propelled particles that do not feedback on the flow, and the effects of activity on flow are far less explored.

In this contribution, we experimentally investigate how swimming bacteria affect the transport and mixing of a passive scalar in a time-periodic chaotic flow. Dye experiments show that bacteria can significantly hinder large-scale transport and global mixing rates. These results are further characterized by computing the stretching fields from experimentally measured velocity fields, which show that bacteria attenuates large-amplitude fluid stretching. This leads to a lower mean finite-time Lyapunov exponent, indicating that activity also decreases flow chaoticity. At small scales, however, bacteria activity can substantially increase local mixing. These two effects lead to a new balance in the dynamical system characterized by a delay in the formation of persistent structures that are overall more homogeneous. A simple numerical simulation reveals the potential mechanism for the experimental phenomena being the transient accumulation of swimming particles along the unstable manifolds of the flow.  

%%%%% Methods and Dye Mixing Experiments
We use the flow cell setup \cite{Voth2002,Voth2003} to create a two-dimensional mixing flow with time-periodic magnetic forcing (Fig. 1\textit{A}; see \textit{SI Appendix} for details). As a sinusoidal voltage is imposed, the induced Lorentz forces in the fluid create a vortex array of alternating vorticity (Fig. 1\textit{B}). The size of each vortex corresponds to the magnets spacing, $L$ = 6 mm. The flow is characterized by two dimensionless numbers. The first is the Reynolds number, $Re=\bar{U}L/\nu$, where $\bar{U}$ is the root mean square (RMS) velocity, and $\nu$ is the fluid kinematic viscosity. The second is the path length, $p=\bar{U}/Lf$, with $f$ being the driving frequency of the flow, which describes the normalized mean displacement of a typical fluid parcel in one period. To better contrast the effects of bacterial activity, we keep the $Re$ and $p$ at approximately 14.0 and 2.3 in all our experiments, respectively. These conditions ($Re$, $p >1$) are known to produce chaotic mixing \cite{Voth2003, ArratiaJSP_2005}, even though the flow preserves spatial and temporal symmetries.

%%% Bacteria info. and dye mixing 
The effects of activity are investigated by adding swimming \textit{E. coli} to the buffer solution (see \textit{Materials and Methods}). The swimming speed of the bacteria ranges from 10 - 20 $\mu$m/s \cite{Patteson2015}. The bacteria volume fraction $\phi_b$ is adjusted from 0\% (pure buffer) to 0.9\%; experiments with  non-motile bacteria are performed at $\phi_b =0.9\%$. This volume fraction range is considered dilute \cite{Kasyap2014}, without introducing large-scale collective motion. We note that the active P\'eclet number, $Pe=\bar{U}L/D_\text{eff}$, is much larger than unity ($\sim 10^6$), where $D_\text{eff}$ is the effective diffusivity of \textit{E. coli} \cite{Patteson2016}. Passive dye mixing experiments are performed by adding a minute amount of fluorescent dye to the fluid. Initially, the flow cell is partitioned by a solid barrier into two halves, one with and one without dye (Fig. 1\textit{C}). The barrier is then lifted and dye penetrates to the undyed portion with time or number of cycles $N$. The duration of each experiment is approximately 30 minutes or $N\approx 400$, over which the bacterial motility remains roughly constant (See \textit{SI Appendix}, Fig. S11). We also perform particle tracking velocimetry (PTV) to obtain the velocity fields of the flow (Fig. 1\textit{B}; see \textit{Materials and Methods}). 

%%%%%%%%%% Results and Discussion
An example of the effects of the bacteria on dye mixing is shown in Fig. 1\textit{C}, for pure buffer ($\phi_b=0\%$) and an active suspension ($\phi_b=0.9\%$). The snapshots are taken at $N=300$ (see Movie S1 - S2 and \textit{SI Appendix}, Fig. S1 for other times and non-motile data). A comparison between the two snapshots  reveals that, in the presence of motile bacteria, the dye penetrates to the undyed region much slower than the buffer and non-motile (Fig. S1) cases. This indicates that bacterial activity is hindering large-scale (dye) mixing and transport. A closer look at the images show that finer dye structures within a vortex cells are also modified when motile bacteria is present; bacterial activity leads to smoother and less structured concentration fields. Normalized concentration gradients fields (Fig. 1\textit{D}) show that, for both cases, gradients are steeper near the flow separatrices (see Fig S3 for non-motile case). This suggests the passive scalar is transported by regions of highest flow strain or material stretching \cite{Voth2002, Arratia2006}. However, the gradient fields in the active case ($\phi_b=0.9\%$) are broader and coarser, suggesting a higher diffusion with bacterial activity. These observations are consistent for all bacterial volume fractions $\phi_b$ investigated here (see \textit{SI Appendix}, Fig. S2 for other $\phi_b$ and non-motile data). These experiments indicate that the presence of swimming bacteria affect both large-scale transport and small-scale mixing of the passive scalar in the time-periodic flow.

The overall mixing rate can be characterized by the variance of dye concentration field, $\langle C^2\rangle$, where $\langle\cdot\rangle$ denotes the spatial ensemble average. We find an exponential decay of the normalized $\langle C^2\rangle$ with time or $N$ for all cases (Fig. 2A), a behavior that is consistent with observations in (time-periodic) chaotic flows \cite{Rothstein1999,Voth2003}; solid lines are exponential fits to the data, $C_0\exp(-RN)$, for $20 < N < 250$. We note that the decay is slower as $\phi_b$ is increased, which is quantified by a linear decrease in mixing rate $R$ with $\phi_b$ (Fig. 2\textit{A}, \textit{Inset}). We expect the linearity to breakdown for non-dilute bacterial suspension due to hydrodynamic interaction and collective behavior. The mixing rate $R$ for the non-motile case is nearly
%%%%%%%% Figure 1
\begin{figure}[H]
\centering
\includegraphics[width=3.412in]{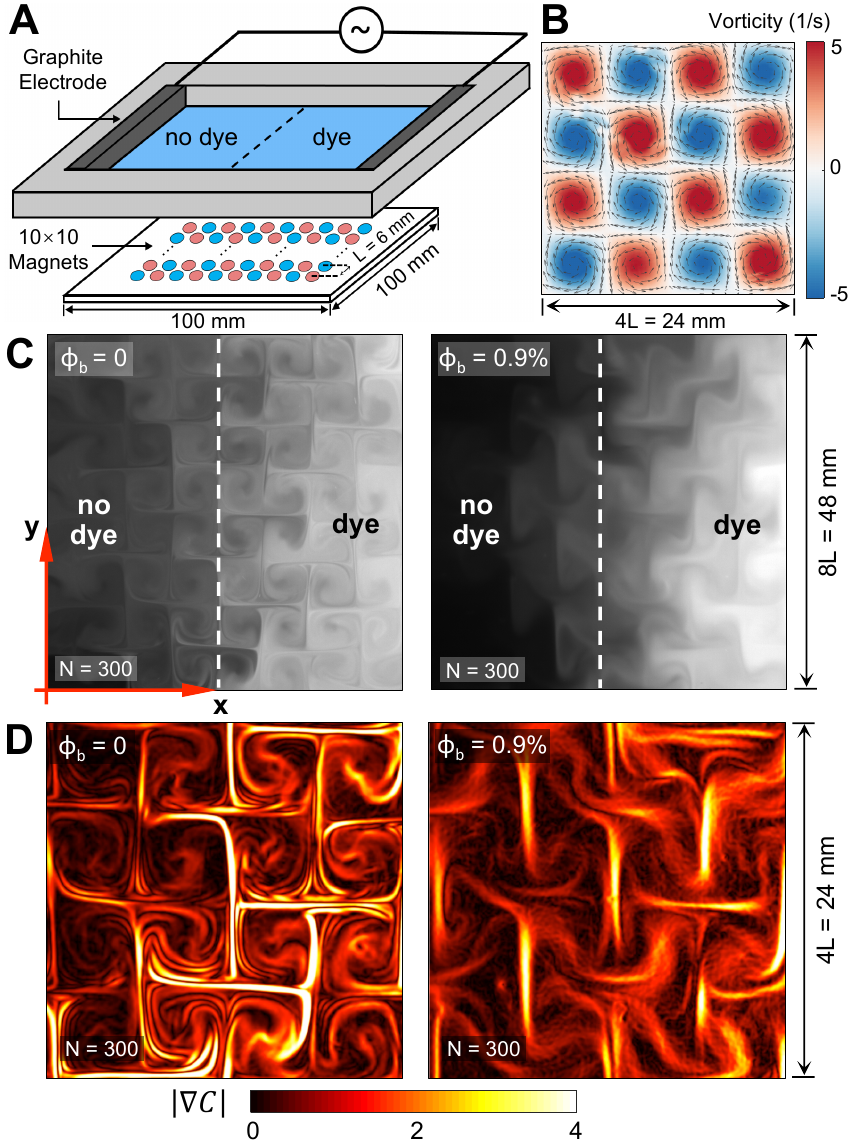}
% \vspace{-6mm}
\caption{
(\textit{A}) Schematic of the flow cell apparatus. A thin layer of buffer solution with \textit{E. coli} is placed above an array of magnets of alternating polarities (denoted by difference colors). A sinusoidal current induces Lorentz force in the fluid to drive the periodic mixing. The right half of the fluid is labelled with fluorescent dye. (\textit{B}) Vorticity field (colormap) and velocity field (arrows) of the periodic mixing flow, corresponding to the first peak value in a time period (5 seconds). The data is measured in a region of 24 $\times$ 24 mm${}^\text{2}$, at a Reynolds number of \textit{Re} = 16.8. (\textit{C}) Photographs of dye mixing experiments in a 48 $\times$ 48 mm${}^\text{2}$ region, for different bacteria volume fractions: $\phi_b = 0\%$ (left), and $\phi_b=0.9\%$ (right). The data is taken after \textit{N} = 300 periods of mixing at a frequency \textit{f} = 0.2 Hz and Reynolds number \textit{Re} = 16.8. (\textit{D}) The magnitude of concentration gradient of the dye field in Fig. 1\textit{C}, for $\phi_b = 0\%$ (top), and $\phi_b=0.9\%$ (bottom), enlarged to a region of 24 $\times$ 24 mm${}^\text{2}$ for better illustration. The data is normalized by the root mean square (RMS) concentration gradient.
}
\label{fig1}
\end{figure}
\noindent double that of the active case at the same $\phi_b$. This indicates that the observed decrease in large scale mixing is not simply a viscous effect, since the addition of passive rod-like particles to the fluid will only lead to an increase in viscosity of 3\% (see \textit{SI Appendix}). We also observe the mixing slows down and deviate from the exponential decay at later times ($N>300$). This slower decay mode or mixing rate is governed by longest wavelength in the flow \cite{Voth2003}, and it sets in earlier for active suspensions. These results show that bacteria activity hinders large-scale transport and decreases the global mixing rates of a passive scalar in time-periodic flows.

%%%%%%%% Figure 2
\begin{figure*}[t!]
\centering
\includegraphics[width=7in]{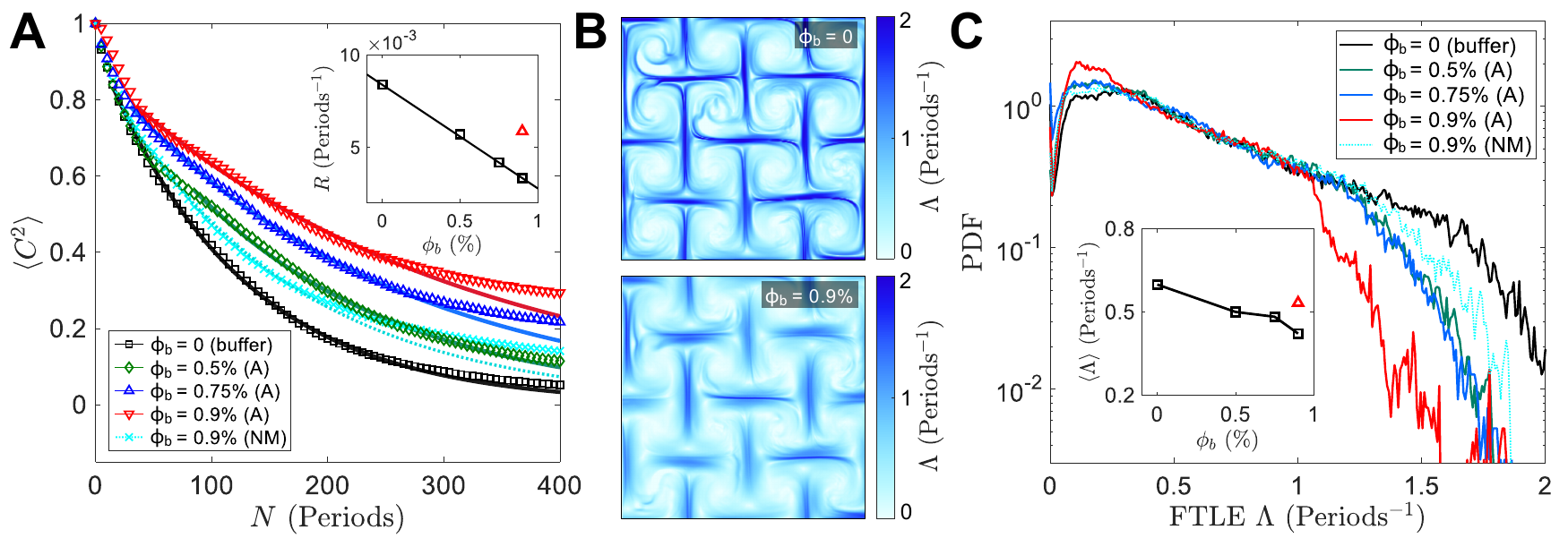}
\caption{(\textit{A}) Decay of the normalized scalar variance $\langle C^2\rangle$ as a function of time (in the unit of periods), for the buffer ($\phi_b = 0$), active suspension (A), and non-motile bacteria suspension (NM). All data is normalized to have an initial condition of unity. Solid curves are exponential fits to the data, $\langle C^2\rangle\sim C_0\exp(-RN)$. \textit{Inset} shows the mixing rate $R$, which decreases linearly with bacterial volume fraction $\phi_b$. The errors in the estimation of the mixing rate $R$ from the exponential fits are less than $2.0\%$. (\textit{B}) The backward (time) finite-time Lyapunov exponent (FTLE) fields, related to the past stretching by $\Lambda=(\log S)/\Delta t$, for $\Delta t = 1.5\ \mathrm{Periods}$. Top: buffer solution ($\phi_b= 0$); bottom: active suspension ($\phi_b= 0.9\%$). The magnitude of $\Lambda$ is attenuated in the presence of bacteria, while the shape of the structures remains similar. (\textit{C}) Probability distributions of the backward FTLE, showing attenuation at high amplitudes ($\ge1\ \mathrm{ Periods}^{-1}$) with increasing $\phi_b$. \textit{Inset} displays a decrease in mean FTLE value $\langle\Lambda\rangle$ as a function of $\phi_b$; non-motile bacteria suspension has a much higher $\langle\Lambda\rangle$ value compared to active suspension of the same $\phi_b$.
}
\label{fig2}
\end{figure*}

%%% Stretching fields (FTLE)
A key property of flows exhibiting chaotic advection is the exponential divergence of nearby trajectories in real space, usually characterized by the largest finite-time Lyapunov exponent (FTLE) $\Lambda$ over a time interval $\Delta t$. One can relate $\Lambda$ to the stretching experienced by a fluid parcel by considering the deformation of an infinitesimal circular fluid element initially located at $(x,y)$. The stretching $S$ is defined as the ratio of the final major diameter (after $\Delta t$) to its initial diameter, and $\Lambda=(\log S)/\Delta t$. Here, stretching fields $S(x,y)$ are computed from experimentally measured velocity fields (see \textit{SI Appendix} for details). Two different quantities are computed at each point, namely past and future stretching. These quantities, past and future stretching, tend to be large on the unstable and stable manifolds of the hyperbolic fixed points of the Poincar\'e map, respectively \cite{Haller2000}. The past stretching fields for both the buffer and active suspension ($\phi_b=0.9\%$) are highly heterogeneous, being much larger along the flow separatrices near hyperbolic points (Fig. 2\textit{B}, see Fig. S5 for non-motile case). While the structure of the field are relatively similar for all cases, stretching is clearly attenuated for the active case, particularly at regions of large stretching. Indeed, the probability distribution of stretching shows that bacteria systematically suppress large values of stretching (Fig. 2\textit{C}). Note that the probability function shifts and decays at lower stretching values (than buffer and non-motile cases) as the concentration of motile bacteria is increased. This behavior is captured by computing the spatially averaged FTLE, which decreases as bacterial concentration increases (Fig. 2\textit{C}, \textit{Inset}). Similar to the mixing rate results, the non-motile case shows larger value of $\langle\Lambda\rangle$ compare to active case at same $\phi_b$. Overall, these results show that bacteria hinders large scale transport and mixing by suppressing the stretching of fluid elements.

% About spectra
The effects of bacteria activity across different length scales are examined by the spatial power spectrum of dye concentration field, $E_c(k)$, (see \textit{SI Appendix} for details). The spectra $E_c(k)$ characterize the fluctuation of the scalar field across wavenumbers \cite{batchelor1959, Williams1997}, whose total spectral power is the scalar variance $\langle C^2\rangle=\int_0^\infty E_c(k)\,dk$. An example of $E_c(k)$ for the buffer, non-motile, and active cases measured at $N=300$ is shown in Fig. 3\textit{A} (see \textit{SI Appendix}, Fig. S8 for $E_c(k)$ at other times). We observe a power law of $E_c(k)\sim k^{-2.0}$ that spans a substantial range of $E_c(k)$, including wavenumbers above $k_L$ and below $k_\eta$. Here, $k_L=2\pi/L$ is the energy injection scale, as shown by peaks in the spectra, and $k_\eta=2\pi/\eta$ is the viscous cutoff scale \cite{Kolmogorov1941}, estimated by setting the local Reynolds number $Re_\eta=u_\eta\eta/\nu\sim1$. We find that the spectral power increases nearly uniformly with $\phi_b$ in the range of $k_L<k<k_\eta$, while the change in logarithmic slope is more apparent at smaller scales, $k>k_\eta$. Note that the area under $E_c(k)$ is larger for active suspensions, which is consistent with a larger remaining scalar variance (Fig. 2\textit{A}). The increases in the total spectral power with $\phi_b$ suggest the rate of transfer of scalar variance from large scale (low $k$) to small scale (high $k$) is hindered by bacterial activity compared to the buffer and non-motile cases. 

% About Correlation
We now focus on large-scale structures in the scalar field by computing the spatial autocorrelation function, defined as:
%%%%%%%% Figure 3
\begin{figure}[H]
\centering
\includegraphics[width=3.412in]{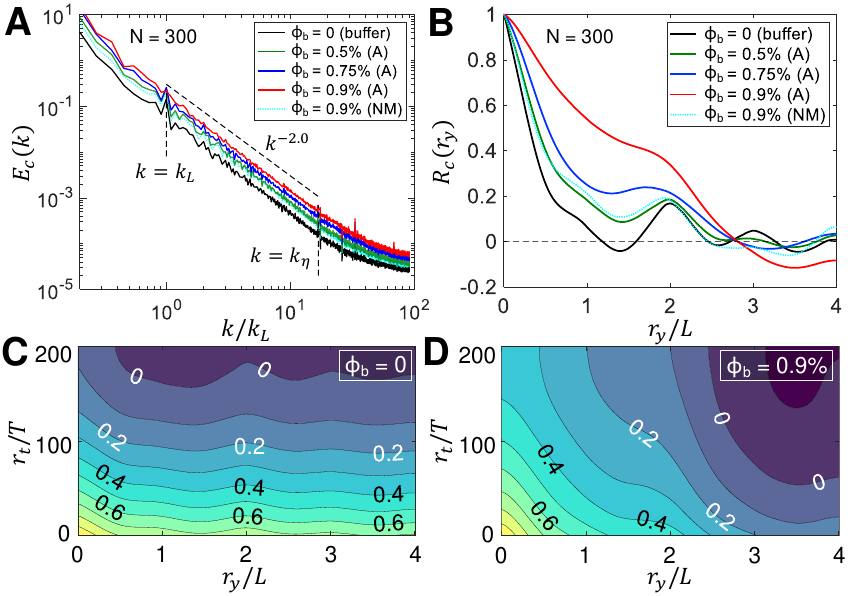}
\caption{(\textit{A}) Concentration power spectra, $E_c(k)$, measured at \textit{N} = 300 for the buffer, active suspension (A) and non-motile suspension (NM). The inclined dashed line indicates a power law of $E_c(k)\sim k^{-2.0}$. Two vertical dashed lines mark the energy injection scale $k_L$, and the viscous cutoff scale $k_\eta$. (\textit{B}) Spatial autocorrelation functions $R_c(r_y)$ of the dye concentration, measured at \textit{N} = 300 for different $\phi_b$.  (\textit{C},\textit{D}) Contour plots of the spatial-temporal autocorrelation function $R_c(r_y,r_t)$, for (\textit{C}) the buffer of $\phi_b=0$, and (\textit{D}) active suspension of $\phi_b=0.9\%$. 
}
\label{fig3}
\end{figure}
\noindent $R_c(r_y)=\langle C(y)C(y+r_y)\rangle/\langle C^2\rangle$, where $y$ is the direction normal to the mean scalar gradient set by the initial condition (see Fig. 1\textit{C}). The correlation $R_c(r_y)$ is plotted for different $\phi_b$ as well as buffer and non-motile cases in Fig. 3\textit{B}, which shows (bacterial) activity leads to increasingly stronger spatial correlations, especially at $r_y<2L$ (within 2 vortex cells). The increase in the spatial correlation indicates the persistence of larger structures in the concentration field even after relatively long time at $N=300$. This result suggests the stretching and folding mechanism responsible for creating thinner striations is hindered by bacterial activity, consistent with stretching field measurements. While we find long-range correlations and spatial symmetries in the scalar fields in the buffer and non-motile cases (peaks at $r_y=2L$, $4L$), these are nearly erased by bacterial activity ($\phi_{b}=0.9\%$). The absence of such peaks suggests that spatial periodicity of large-scale structures is broken by the bacterial activity.

Next, we examine the time evolution of dye structures by computing the spatial-temporal autocorrelation function $R_c(r_y,r_t)=\langle C(y,t)C(y+r_y,t+r_t)\rangle/\langle C^2\rangle$. For the buffer case, the contour plots of $R_c(r_y,r_t)$ quickly develop an invariant sh-
%%%%%%%% Figure 4
\begin{figure}[H]
\centering
\includegraphics[width=3.412in]{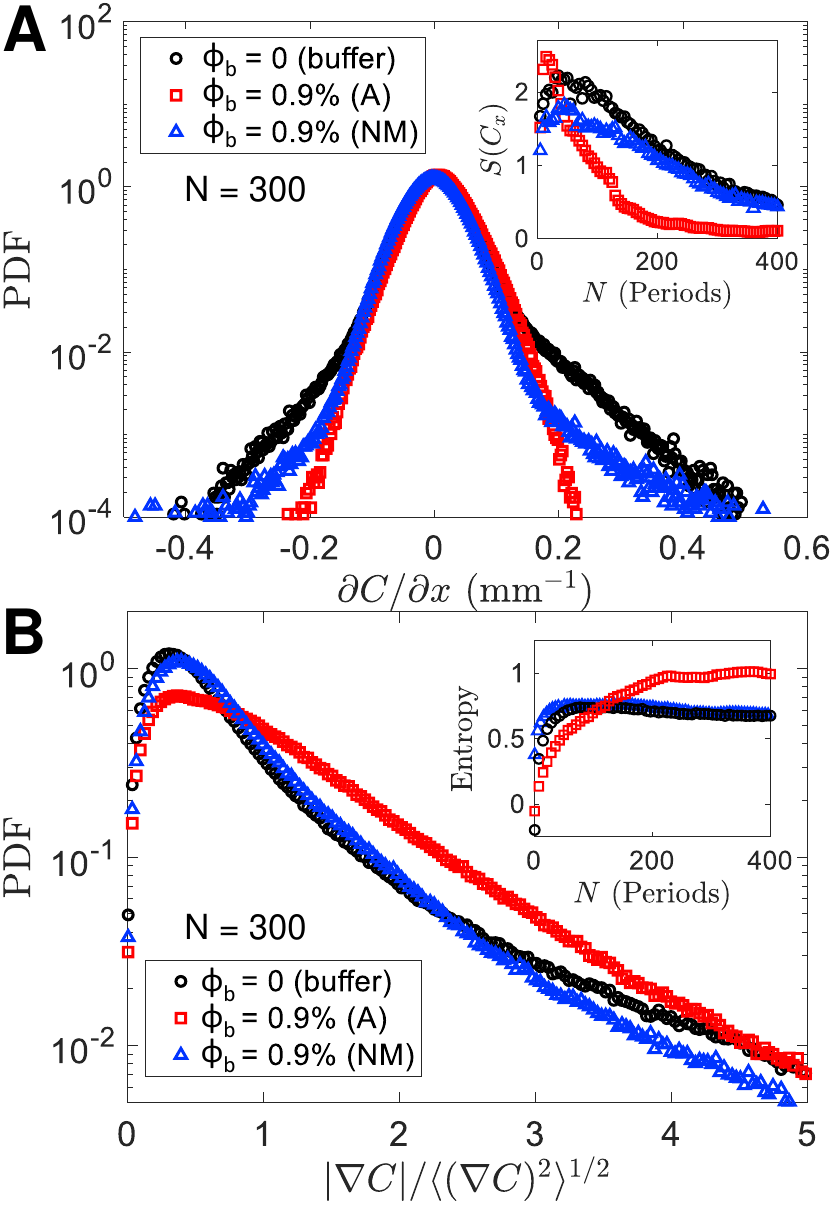}
% \vspace{-1mm}
\caption{(\textit{A}) Probability distribution of the partial gradient $\partial C/\partial x$, measured at \textit{N} = 300 for the buffer ($\phi_b=0$), active bacteria suspension ($\phi_b=0.9\%$, A), and non-motile bacteria suspension ($\phi_b=0.9\%$, NM). \textit{Inset} shows the skewness of these PDFs as a function of time. (\textit{B}) Probability distribution of the concentration gradient magnitude $|\nabla C|$, normalized the root mean square (RMS) gradient $\langle (\nabla C)^2\rangle^{1/2}$, measured at \textit{N} = 300 for the buffer, active suspension (A), and non-motile bacteria suspension (NM). \textit{Inset} shows the differential entropy of the same PDFs versus time.}
\label{fig4}
\end{figure}
\noindent ape, and the decay of the correlation with time is nearly uniform in space for all $r_y$ (Fig. 3\textit{C}). The initially formed invariant structures become decorrelated for $N> 150$ due to diffusion. However, for the $\phi_b=0.9\%$ case, we find that structures decorrelate much slower in time; correlations in the range $r_y<2L$ remain relatively high up to 200 periods. We believe that, for $r_y>2L$ (larger than 2 vortex cells), the structures are decorrelated due to local stochastic behavior of bacteria activity in each vortex cell (this will be discussed in more detail later). This slow decay in correlation shows the presence of larger structures in the concentration field, which are less susceptible to diffusion. Taken together, these results show that bacterial activity leads to a decrease in long-range spatial correlation but an increase in temporal correlation. The structures created by the flow of these active fluids are coarser and more long-lasting than the passive case. 

%%%% small-scale mixing
So far we have shown that bacterial activity can hinder large-scale transport and mixing by hindering the production of fine structures that decorrelate faster in time. To gain insights into mixing small scales, we examine the scalar gradient fields (Fig. 1\textit{D}). We further quantify the results by computing the probability density function (PDF) of the partial scalar gradient $\partial C/\partial x$ (Fig. 4\textit{A}), whose skewness is an indicator of the local isotropy of scalar fields \cite{Sreenivasan_PRSL_1991}. Here, $x$ is defined as the direction of the mean scalar gradient (see Fig. 1\textit{C}). These PDFs have a non-Gaussian core with exponential tails at high gradient values. Remarkably, we find the PDF for active suspension ($\phi_b=0.9\%$) is nearly symmetric, while the PDFs for the buffer and non-motile bacteria suspension are asymmetric. Further calculations show the skewness values in the active suspension decrease by an order of magnitude relative to the buffer and non-motile cases (Fig. 4\textit{A}, \textit{Inset}). The active suspension has a vanishing steady state skewness of 0.09, suggesting that the scalar field is almost isotropically distributed at small scales. This result is unexpected since local isotropy of scalar fields is rarely observed; the skewness of $\partial C/\partial x$ is at the order of unity (rather than zero) even at geophysical Reynolds number \cite{Sreenivasan_PRSL_1991,Warhaft2000}. Stronger mixing at the ``swimmer'' scale was previously proposed \cite{Wu_PRL_2000,Kurtuldu_PNAS_2011,Poon_PRE_2013}, mostly by observing an increase in particle mean-square-displacement (MSD) and effective diffusivity in quiescent flows. Here, in the presence of an imposed flow, we show that the swimmer-flow interaction can even lead to local isotropy in the scalar field.

%%%%%%%% Figure 5
\begin{figure*}[t!]
\centering
\includegraphics[width=4.5in]{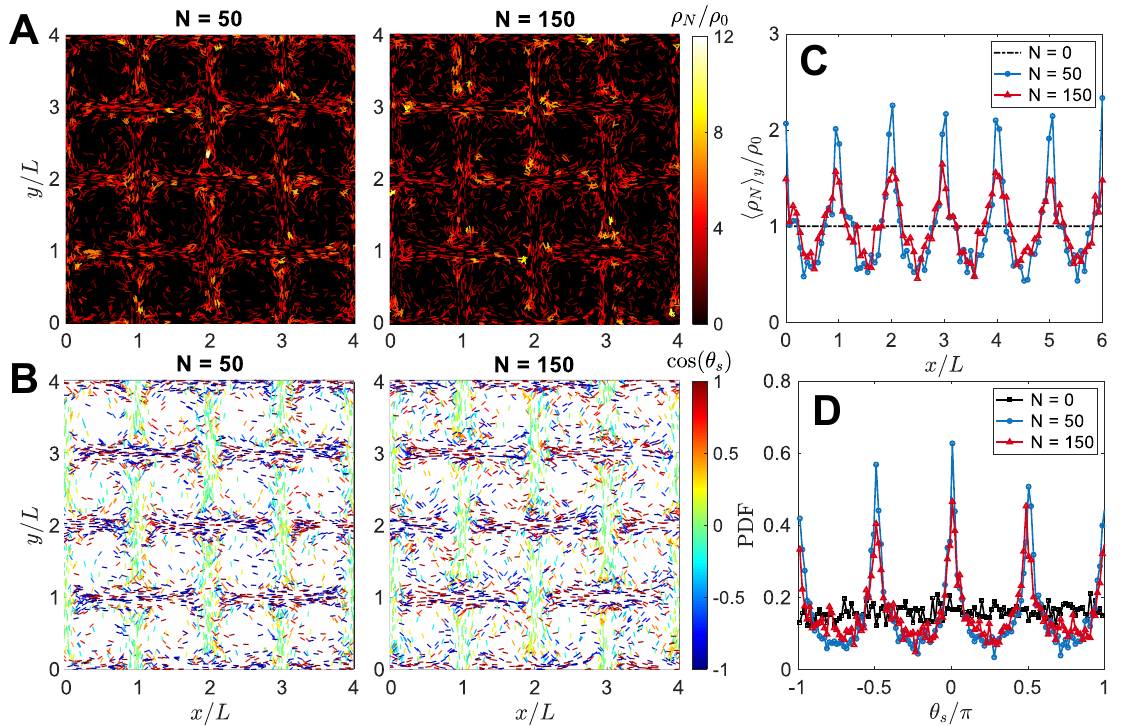}
% \vspace{-2mm}
\caption{Numerical simulations of the transport of swimming particles. (\textit{A, B}) Particle distribution in time-periodic flows of \textit{Re} = 14.0, \textit{p} = 1.94; snapshots are taken at \textit{N} = 50 (left) and \textit{N} = 150 (right). Swimming particles are visualized as rods that are colored by (\textit{A}) particle number density $\rho_N$ normalized by initial density $\rho_0$ and (\textit{B}) the cosine of particle swimming angles $\cos(\theta_s)$, respectively. (\textit{C}) Horizontal profile of the vertically averaged number density $\langle\rho_N\rangle_y$ normalized by $\rho_0$, showing accumulation of particles at the locations of the multiple of $L$. The accumulation becomes less intense at longer times. (\textit{D}) The probability distribution of the swimming angle $\theta_s$ of all particles, indicates the alignment of particles with the vertical and horizontal unstable manifolds by showing peaks at the multiple of $\pi/2$. The alignment becomes less strong at later times.
}
\label{fig5}
\end{figure*}

The effectiveness of the mixing process at the finer scales can be characterized by the PDF of the magnitude of scalar gradient $|\nabla C|$, normalized by $\langle (\nabla C)^2\rangle^{1/2}$ to compensate for the decay of contrast (Fig. 4\textit{B}). These PDFs reach invariant forms with time (see \textit{SI Appendix}, Fig. S9) that have a nonzero mean and a notably exponential tail. The invariant form is also known as ``persistent pattern'' that is typical for time-periodic mixing \cite{Rothstein1999}. And exponential distribution at high gradients are expected for a passive scalar subjected to a mean gradient undergoing random advection \cite{Jayesh1991, Holzer1994, Pumir1991}. Here, however, we show that the active suspension ($\phi_b=0.9\%$) reaches a distinct invariant form than the buffer and non-motile cases, characterized by a longer and more pronounced exponential tail. It suggests that bacteria activity may be enhancing the randomness of the local advection of dye. Moreover, bacteria can also delay the formation of the invariant form by as much as 100 periods relative to the buffer and non-motile cases, which is captured by the differential entropy of the PDFs (Fig. 4\textit{B}, \textit{Inset}, see \textit{SI Appendix} for the calculation). This is because stretching is adversely affected by the activity. The balance between these two effects, i.e. the decrease in stretching and the enhancement in diffusion, leads to a slower yet better small-scale mixing, as shown by a larger entropy in the active suspension (Fig. 4\textit{B}, \textit{Inset}, $N>200$).

%%%%%% Simulation
To gain further insight into these results, we perform numerical simulations of swimming particles in the flow (See \textit{Materials and Methods} for details). Initially, swimmers are uniformly distributed in the simulation domain with random orientations. As the simulation begins, we see strong accumulation of particles along the flow unstable manifolds (Fig. 2\textit{B}) at an early time of $N=50$; this is also shown by the number density pattern in Fig. 5\textit{A} (see Movie S3 for the transient). The local particle number density $\rho_N$ at $N=50$ is an order of magnitude larger than the initial density $\rho_0$. At a later time of $N=150$, the particle accumulation reduces as the number density patterns grow wider and more disperse (Fig. 5\textit{A}, right panel). This transient behavior is further quantified by the one-dimensional number density profile (Fig. 5\textit{C}). Swimming particles, initially evenly distributed, form strong peaks at each multiple of $L$ where unstable manifolds are located. These peaks becomes weaker when particle distribution becomes more disperse at later times.

Simulation results also show that swimmers align with the unstable manifolds as characterized by the cosine of their swimming angle $\theta_s$, as shown in Fig. 5\textit{B} (see \textit{SI Appendix}, Fig. S11 for $\sin\theta_s$). Nearly all particles accumulated near vertical unstable manifolds are swimming upward or downward (green, $\cos\theta_s=0$); and almost all particles near horizontal manifolds are swimming leftward or rightward (blue/red, $\cos\theta_s=\pm1$). The alignment of particles become weaker at later times ($N=150$) as the particle distribution broadens (Fig. 5\textit{B}, right). This is characterized by the probability distribution of the swimming angle (Fig. 5\textit{D}). While initially swimming angle is uniformly distributed, at later times we find a decay in all peaks as more particles fall between intermediate angles. While the steady state  accumulation of active particles along unstable manifolds have been previously observed \cite{Khurana2012,Guasto_PNAS2019}, here, however, we show that this accumulation may be transient and non-monotonic in time (see \textit{SI Appendix}, Fig. S12). Taken together, these simulation results suggest that the accumulation and alignment of swimmers along the unstable manifolds may be responsible for the decrease in fluid stretching. We believe that this accumulation may lead to an increase in local extensional viscosity or resistance of stretching along the manifolds, leading to a decrease in local FTLE (Fig. 2\textit{C}). This causes a decrease in mixing rate (Fig. 2\textit{A}) and an increase in average structure size (Fig. 3\textit{B}). 

We can estimate the increase in extensional viscosity ($\mu_{\varepsilon}$) due to local bacteria accumulation. For dilute suspensions of rod-like pusher swimmers, the extensional viscosity in the high P\'eclet number limit can be estimated as $\mu_{\varepsilon}=3 \mu_{0}\left(1+\phi_b \alpha^{2} \pi /(18 \ln 2 \alpha)\right)$, where $\mu_0$ is the shear viscosity of the suspending fluid, and $\alpha$ is the particle aspect ratio \cite{Saintillan_PRE_2010} (see \textit{SI Appendix} for details). Using this expression, we estimate that non-motile bacteria suspension of $\phi_b=0.9\%$ will only lead to a uniform increase of 2.7\% in extensional viscosity. However, an active suspension of same $\phi_b$ has a much higher local volume fraction ($\phi_b=10.8\%$) due to bacteria accumulation near the manifolds, leading to an increase in $\mu_{\varepsilon}$ of 33.3\% (See \textit{SI Appendix}). We believe that this relatively large increase in local extensional viscosity along the unstable manifolds is responsible for attenuation of stretching and the observed hindrance in large-scale mixing. 

%%%%%%% Conclusion
Fluid mixing is an important phenomenon that occurs in diverse natural situations (e.g., lakes, rivers, atmosphere). Here, we explored how swimming microorganisms affect mixing and transport in flows exhibiting chaotic advection. We find that swimming \textit{E. coli} can hinder large-scale transport of a passive scalar (i.e. dye) in time-periodic flows even though small-scale (local) mixing is enhanced. Bacteria activity can adversely affect large-scale mixing by suppressing the stretching and folding mechanism, a hallmark of chaotic advection. This was demonstrated explicitly by measuring the flow stretching fields; regions of large stretching were attenuated by the presence of bacteria, which resulted in a lower mean of finite-time Lyapunov exponent (FTLE) as bacteria concentration is increased. In other words, bacteria activity reduces the chaoticity of the flow. The attenuation of stretching results in coarser and lasting structures, as shown by the increase in both spatial and temporal autocorrelations with swimming \textit{E. coli}. This is in contrast to the strong local mixing produced by bacteria at later times, which can even lead to nearly locally isotropic dye concentration fields. This strong local mixing coupled with attenuation in stretching leads to a dynamical system in which scalar field invariant behavior is delayed; this invariant field is, however, subjected to higher randomness at the finer scales for $\phi_b >0$, as shown by the longer exponential tail in Fig. 4\textit{B}. Simulation results suggest that the attenuation of stretching and mixing at early times may come from an increase in local extensional viscosity due to the accumulation of swimmers along unstable manifolds. Overall, our results provide new insights into how the nonlinear coupling between flow and activity work together to transport and homogenize scalars such as impurities and temperature. 

\matmethods{\sffamily
\subsection*{Bacterial Suspension Preparation}
We inoculated a strain of \textit{Escherichia coli} (wild type K12 MG1655) into LB broth (Sigma-Aldrich) liquid media. The media were then incubated at 37 $^\circ$C and 135 rpm overnight (12 - 14 hours) to attain a stationary phase of approximately 10$^9$ cell/mL cell density \cite{Patteson2015}. We centrifuged the stationary-phase culture at 5000 rpm for 3.5 min, and re-suspended the pellet into 20 mL of buffer solution. The buffer is an aqueous solution of 2 wt\% KCl, 1 wt\% NaCl, which does not inhibit cell viability \cite{Gandhi2014}. The bacterial volume fraction $\phi_b$ was adjusted from 0.5\% to 0.9\% within the same buffer. Non-motile bacteria suspension was made by killing the \textit{E. coli} with 6\% NaClO solution (1:20 to the culture), and then repeating the aforementioned centrifuging-resuspending protocol. 

\subsection*{Dye Mixing Experiments}
We labelled half of the fluids with 250 $\mu$L of dye (2.5 $\times$ 10$^{ \text{-3}}$ M fluorescein sodium aqueous solution). The dye was then stirred and dispersed uniformly (6.25 $\times$ 10$^{ \text{-5}}$ M) in the labelled fluid, during which a barrier placed in the middle prevented the labelled and unlabelled fluid from mixing. After the barrier was lifted, we imposed a AC voltage of 4 V and 0.2 Hz to the fluid layer to drive the mixing. Images were taken in the (6 cm)$^\text{2}$ center region using a CMOS camera (IO industries, Flare 4M180), operating at 5 fps and 2000$^\text{2}$ pixel resolution. Dye field was illuminated under black light (USHIO, F8T5/BLB). The black light (peak emission 368 nm) is in the ultraviolet (UV) range, but 90\% of energy is in the range of long-wave UVA-I (340 - 400 nm), which does not harm the cell \cite{Vermeulen2008}. A filter (TIFFEN, Yellow 12) was used to cut off UV, such that the light intensity is linearly proportional to dye concentration.

\subsection*{Particle Tracking Velocimetry (PTV)}
We dispersed 100 $\mu$m polyethylene (PE) fluorescent particles in the bacterial suspension. Particle positions were recorded by the aforementioned camera, this time operating at 30 fps and 1140$^\text{2}$ pixel resolution in a (3.6 cm)$^\text{2}$ region. By using particle-tracking software \cite{Crocker1996}, we extracted particle trajectories and measured particle velocities from 6$^{\text{th}}$ order polynomial fitting. The velocities were then phase averaged and interpolated on a spatial grid to obtain the velocity map.

\subsection*{Simulations}
The swimmers are modelled as non-interacting axisymmetric ellipsoids with a swimming speed $v_s$ in the direction of $\mathbf{n}$. The position of the swimmers $\mathbf{x}$ are governed by:
\begin{equation}\label{eqn.position}
    \dot{\mathbf{x}} = \mathbf{v}_f(\mathbf{x},t) +  v_s \mathbf{n},
\end{equation}
where $\mathbf{v}_f$ is the fluid velocity, and $v_s=20\ \mu$m/s for \textit{E. coli}. The orientations of swimmer can be modelled using Jeffery's dynamics \cite{Jeffery1922}:
\begin{equation}\label{eqn.orientation}
    \dot{\mathbf{n}}=[\mathbf{\Omega}(\mathbf{x}, t)+\gamma \mathbf{D}(\mathbf{x}, t)] \mathbf{n}-\gamma[\mathbf{n} \cdot \mathbf{D}(\mathbf{x}, t) \mathbf{n}] \mathbf{n},
\end{equation}
where $\mathbf{D}$ and $\mathbf{\Omega}$ are the symmetric and skew parts of the velocity gradient tensor $\nabla\mathbf{v}_f$. Here, $\gamma = (1-\alpha^2)/(1+\alpha^2)$ is a shape factor, with $\alpha$ being the swimmer aspect ratio; the factor is roughly $\gamma\approx0.88$ for \textit{E. coli}. The flow in the simulation is defined by a Taylor-Green type stream function:
\begin{equation}\label{eqn.streamfunc}
    \psi = (UL/\pi)\sin(\pi x/L)\sin(\pi y/L)\sin(2\pi ft),
\end{equation}
where $U$ is the maximum velocity, related to RMS velocity by $\bar{U}=U/\sqrt{8}$. The simulation domain is $(6L)^2$ in size, with 10,000 swimming particles in it, and periodical boundary conditions imposed on all boundaries.
}

\showmatmethods{} % Display the Materials and Methods section

\acknow{\sffamily
This work was supported by NSF Grant DMR-1709763. We thank Jeffrey Guasto, Tom Solomon, Kevin Mitchell, and Simon Berman for insightful discussions, and David Gagnon and Madhura Gurjar for the help with early work.
}

\showacknow{} % Display the acknowledgments section
\vspace{+2mm}

% Bibliography
\bibliography{ms}

\end{multicols}

\end{document}

%% Comment out or remove this line before generating final copy for submission; this will also remove the warning re: "Consecutive odd pages found".
% \instructionspage  

\maketitle

%% Adds the main heading for the SI text. Comment out this line if you do not have any supporting information text.
\SItext

\section{The flow cell setup}
The flow cell is an electrolytic cell, which consists of a 10 cm $\times$ 10 cm flow chamber, two graphite electrodes, and an array of permanent magnets beneath it (See Fig. 1\textit{A}). These magnets were placed uniformly with a spacing of $L=6$ mm and alternating polarities. Each magnet has a diameter of 4.8 mm, and a magnetic flux density 0.66 T. We put 25 mL of the electrolyte solution (buffer solution, 3\% wt. KCl and 1\% wt. NaCl aqueous solution) into the flow chamber to form a thin fluid layer. The thickness of the fluid layer in the center region is approximately 2 mm. We then imposed a sinusoidal AC voltage of 4V and 0.2 Hz to the electrolyte via graphite electrodes. The induced Lorentz force of conducting ions in the solution will create spatially-periodic vortices (See Fig. 1\textit{B}). The Reynolds number of the flow were controlled from 2.5 to 20, by adjusting the voltage of the forcing between 0.5 - 4 V. In the experiments of active suspensions, we added swimming \textit{Escherichia coli} (wild type K12 MG1655) into the aforementioned electrolyte (buffer) solutions. The bacterial volume fraction was adjusted between $\phi_b=0-0.9\%$.

\section{Dye mixing experiments}
In the beginning of each experiment, we divided the fluid in the flow chamber into two halves with a physical barrier. We then added 250 $\mu$L of passive dye solution (2.5 $\times$ 10$^{ \text{-3}}$ M fluorescein sodium aqueous solution) to the right half of the chamber. The dye was then stirred and dispersed homogeneously (6.25 $\times$ 10$^{ \text{-5}}$ M) in the labelled fluid. After the barrier was lifted, we electrified the fluid layer with a AC voltage and started the mixing. The passive dye then penetrates from the right to the left half of the chamber. The duration of each experiment is around 30 minutes, or 400 number of periods (each period of 5 s).

In Fig. S1, we show the time evolution of dye concentration field over 300 periods, for the buffer solution ($\phi_b=0$), an active suspension ($\phi_b=0.9\%$), and a non-motile bacteria suspension ($\phi_b=0.9\%$). The dye mixing in active suspensions of other bacterial volume fraction $\phi_b$ are available in Movie S1-S2. We find that the dye solution penetrates to the left much slower in the active suspension (right column), relative to the buffer (left column), and non-motile suspension (middle column). We also find the differences in the shape and intensity of the dye structure. The changes in local dye structures can be better illustrated by the spatial gradients of dye concentration field. 

\section{Calculation of dye concentration and gradient field}
The dye was illuminated under black light (USHIO, F8T5/BLB). The black light has a peak emission of 368 nm, and an emission bandwidth of 340 - 400 nm, which is in the ultraviolet (UV) range. The fluorescent dye (sodium fluorescein) has a peak emission of 512 nm. Since we mounted a filter (TIFFEN, Yellow 12) on the lens, UV was totally filtered out, and the camera only received the light emission from the fluorescent dye. Hence, the light intensity is linearly proportional to dye concentration. After proper subtraction of background image, we can get the dye concentration field $C(\mathbf{x})$, which is forced to have a zero mean, i.e. $\langle C\rangle=0$, where $\langle\cdot\rangle$ denotes the spatial ensemble average. Since $C(\mathbf{x})$ has a zero mean, the spatial ensemble average of $C^2$ value, $\langle C^2\rangle$, is the scalar variance that represents the unmixedness of the dye concentration field.

The calculate the gradient, we first preprocessed the images with a 2D Gaussian filter \cite{Haddad_IEEE_1991} to remove the Gaussian noise of the pixels. The kernel of filter has the following expression:
\begin{equation}
    K(x, y)=\frac{1}{2 \pi \sigma^{2}} \exp(-\frac{x^{2}+y^{2}}{2 \sigma^{2}}).
\end{equation}
For our use, the standard deviation is set to be $\sigma=1$, and the kernel size is $5\times5$ pixel${}^2$. After the preprocessing, the gradient was then calculated between the nearby two pixels using the central difference scheme finite difference method, or simply:
\begin{equation}
    f^\prime(x)=\frac{f(x+\Delta x)-f(x-\Delta x)}{2\Delta x}.
\end{equation}

In Fig. S2, we show examples of the magnitude of concentration gradient field $|\nabla C|$ measured at $N=300$ periods, for the buffer ($\phi_b = 0$), active suspensions of different bacterial volume fraction ($\phi_b = 0.5\%, 0.75\%, 0.9\%$), and non-motile bacterial suspension ($\phi_b=0.9\%$). The gradient data is normalized by the root-mean-square (RMS) gradient $\langle (\nabla C)^2\rangle^{1/2}$. In the buffer case, we find sharp gradients near the flow separatrices, and wavy fine structures inside the vortex cells. For active suspensions, the sharp gradient near the separatrices still remain. However, the fine structures inside the vortex cells become wider and coarser. This result indicates bacteria enhance diffusion of the passive dye, and erase the details in the fine structures. Non-motile bacteria suspension, on the other hand, still preserve large amount of fine structures compared to active suspension of the same $\phi_b$. 

%%%%%%%%% Each figure should be on its own page
%%%%%% Figure 1 
\begin{figure}\label{Fig.S1}
\centering
\includegraphics[width=5.4in]{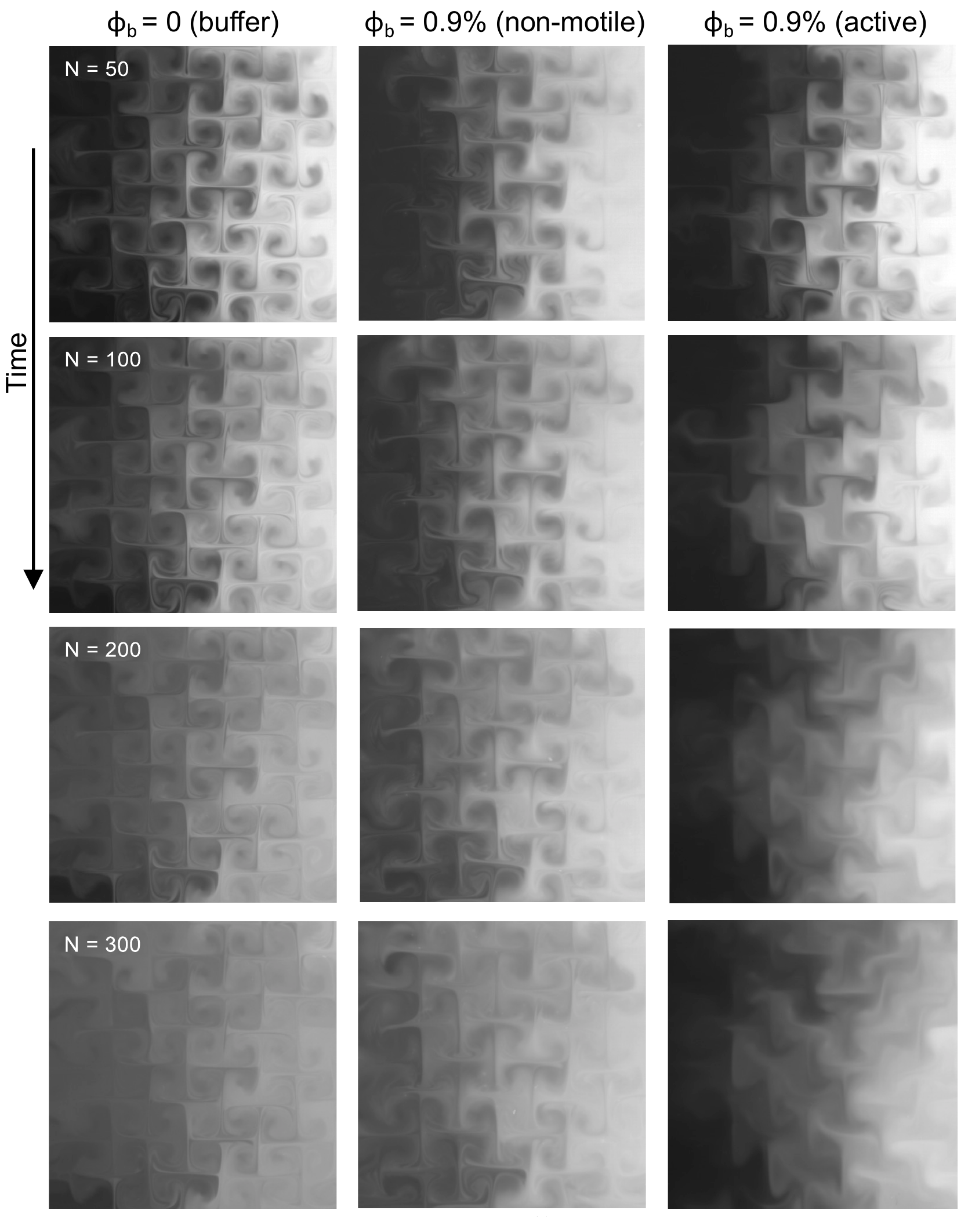}
\caption{Snapshots of the time evolution of the dye mixing patterns, for buffer solution, non-motile bacteria suspension, and active bacteria suspension. Left column: the buffer ($\phi_b=0\%$); middle column: non-motile bacteria suspension ($\phi_b=0.9\%$); right column: active suspension ($\phi_b=0.9\%$). Passive dye is initially confined to the left half of the chamber. We find much slower transport of dye from the right to the left in the active suspension, compared to the buffer solution, and non-motile bacteria suspension. The Reynolds number of the flow is $Re=14.0$ and the path length is $p=2.33$.}
\end{figure}

%%%%%%%%%%%%% Figure 2
\begin{figure}\label{Fig.S2}
\centering
\includegraphics[width=6.2in]{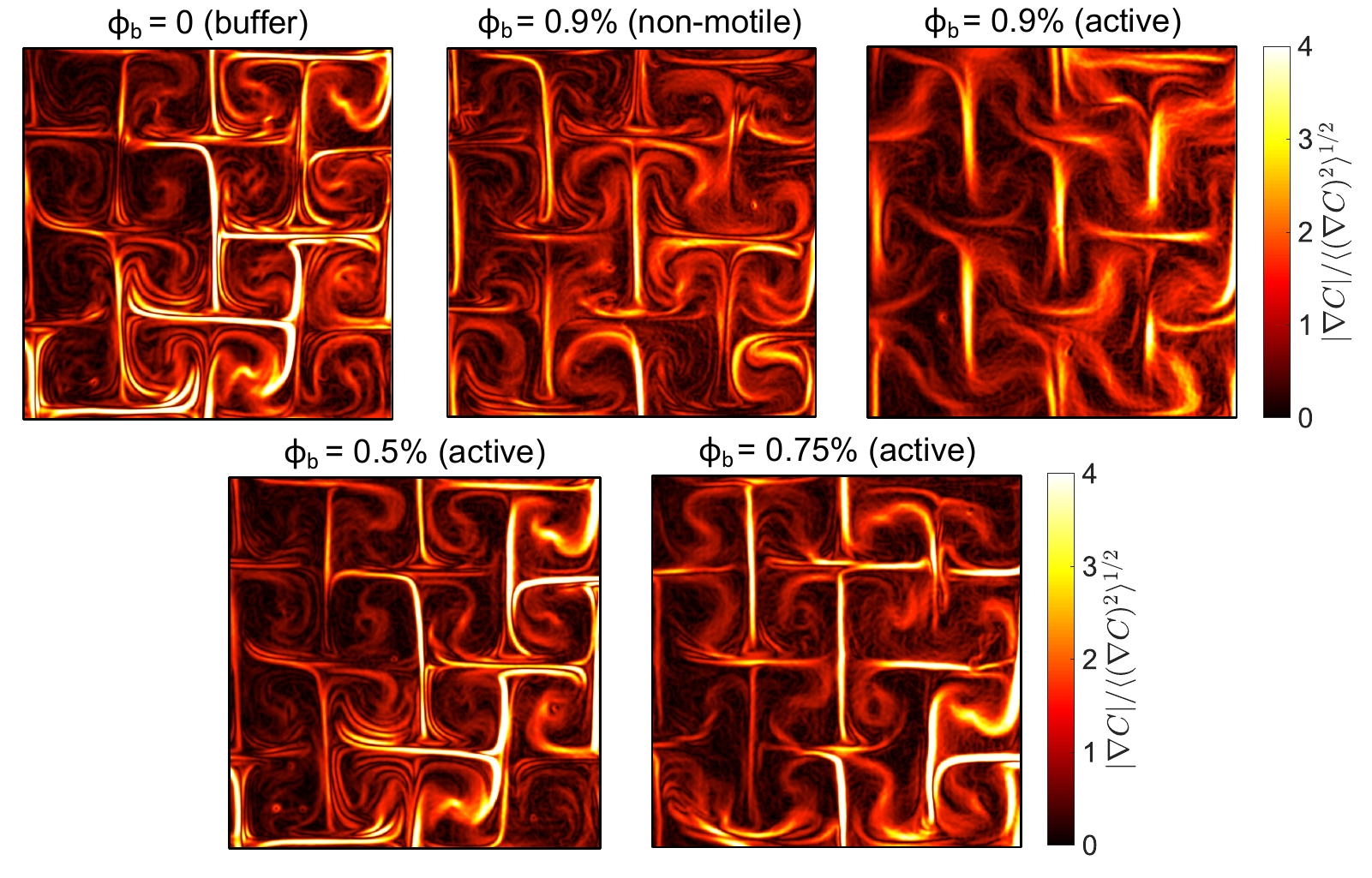}
\caption{The magnitude of dye concentration gradient field $|\nabla C|$, for the buffer, active suspensions of different bacterial volume fraction $\phi_b$, and non-motile bacteria suspension. The data is measured at \textit{N} = 300 periods in a $4L\times4L$ ($24\ \text{mm}\times24\ \text{mm}$) region. The gradient is normalized by the root mean square concentration gradient $\langle(\nabla C)^2\rangle^{1/2}$. We find fine structures in the dye field, marked by sharp gradients, gradually fade away with increasing $\phi_b$ in the active suspensions. Non-motile bacteria suspension, on the other hand, has larger amount of fine structures compared to active suspension of the same $\phi_b$.}
\end{figure}

%%%%%%%%%%%%% Figure 3
\begin{figure}\label{Fig.S3}
\centering
\includegraphics[width=5.4in]{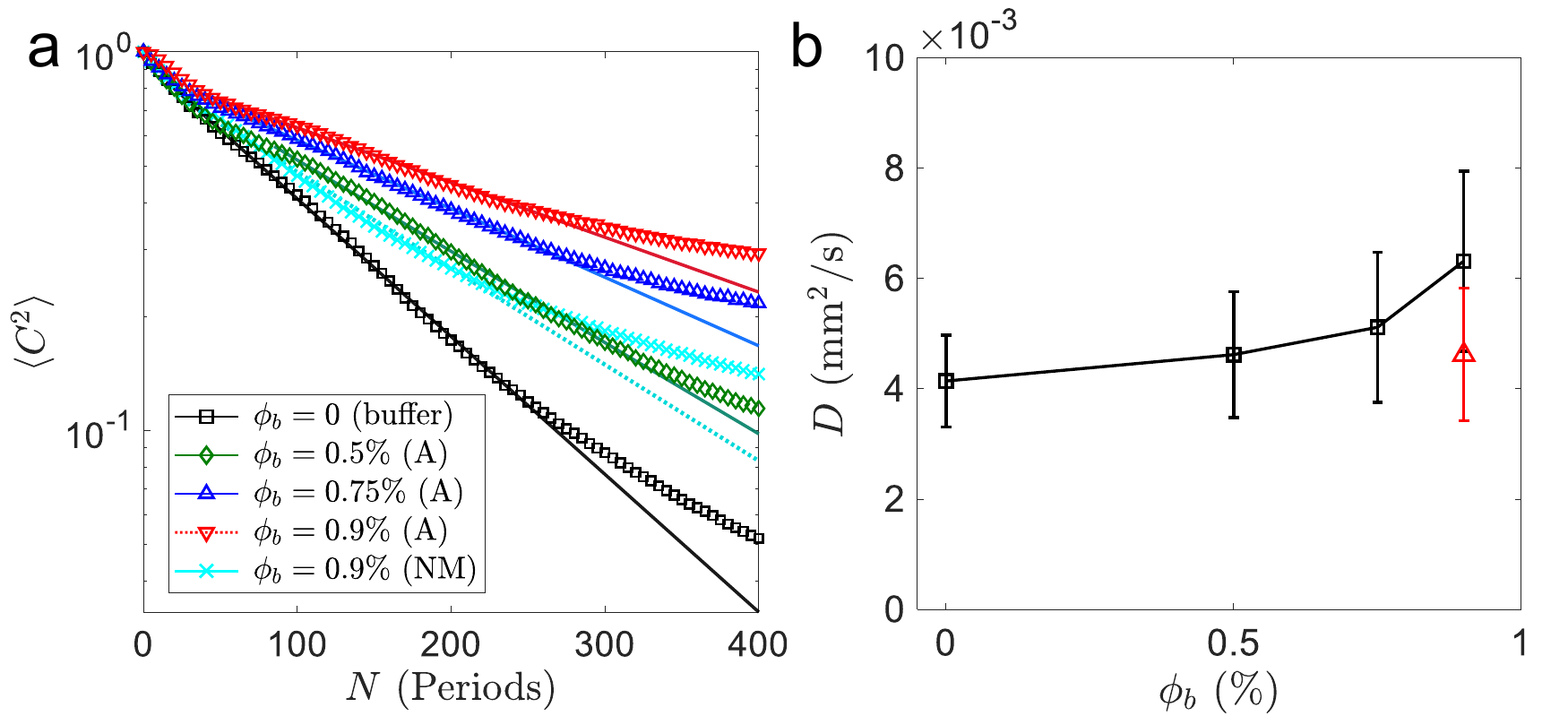}
\vspace{+1mm}
\caption{(\textit{a}) The decay of scalar variance $\langle C^2\rangle$ in logarithmic scale as a function of time (periods), for the buffer ($\phi_b=0\%$), active bacteria suspension (A), and non-motile bacteria suspension (NM). All data is normalized to have a unity initial condition. Solid lines (dashed line for the NM data) are exponential fittings, $C_0\exp(-RN)$, where $R$ is the mixing rate. (\textit{b}) Effective diffusivity of the dye calculated from the decay of scalar variance $\langle C^2\rangle$, which increases nonlinearly with $\phi_b$ for the buffer and active suspensions (black squares). Non-motile bacteria suspension (red triangle) does not seem to increase the effective diffusivity of dye.}
\end{figure}

% (\textit{b}) The mixing rate $R$ from exponential fittings, which decreases linearly as a function of bacterial volume fraction $\phi_b$ for the buffer and active suspensions (green squares). Non-motile bacteria suspension (red star) has a much larger mixing rate compared to active suspension of the same $\phi_b$. 

\section{Dissipation of scalar variance}
We have shown in Fig. 2\textit{A} that the addition of bacteria slows down the decay of the scalar variance $\langle C^2\rangle$, and reduce global mixing rate. And we speculate that these phenomena are due to an increase in local viscosity in the regions of high fluid stretching. On the other hand, non-motile bacteria behave like passive particles that will not accumulate in certain regions of the flow. They will lead to a uniform increase in viscosity in the fluid. 

In Fig. S3a, we shows the decay of variance for the buffer, active suspension, and non-motile bacteria suspension. We find that non-motile bacteria slightly slow down the mixing, because they increase the viscosity of the fluid. But the decay of $\langle C^2\rangle$ in non-motile bacteria suspension is still much faster than the active suspension of the same $\phi_b$. We further calculate the mixing rate $R$ from exponential fittings of the data, $\langle C^2\rangle\sim C_0\exp(-RN)$, as shown by the solid lines in Fig. S3a. The mixing rates are in Fig. 2\textit{A}, \textit{Inset}. The mixing rate for non-motile bacteria suspension of $\phi_b=0.9\%$ is even higher than the active suspension of $\phi_b=0.5\%$. These results suggest that the activity (i.e. swimming motion) of bacteria is essential for us to observe the hindrance phenomena.

% This is further demonstrated by the mixing rate. As shown in Fig. S3b, the mixing rate of non-motile suspension (red star) is much higher than that of the active suspension at the same bacterial volume fraction ($\phi_b=0.9\%$), 

Next, we calculated the effective diffusivity of passive dye in the flow from the decay of scalar variance $\langle C^2\rangle$. The equations we used to estimate the effective diffusivity can be simply derived from advection-diffusion equation of this form:
\begin{equation}\label{advec-diff}
    \frac{\partial C}{\partial t}+(\mathbf{u} \cdot \nabla) C=D \nabla^{2} C.
\end{equation}
We can multiply both sides by $C(\mathbf{x})$ for Eqn. \ref{advec-diff}, and we get:
\begin{equation}\label{step_deriv}
    \frac{\partial}{\partial t}\left[\frac{1}{2} C^{2}\right]+\nabla \cdot\left[\left(\frac{1}{2} C^{2}\right) \mathbf{u}\right]=D\nabla \cdot[C \nabla C]-D(\nabla C)^{2}.
\end{equation}
We can then take the ensemble average $\langle\cdot\rangle$ for both sides. According to the argument of Batchelor \cite{batchelor1959}, the second and third terms of Eqn. \ref{step_deriv} are transport terms that only move $C^2$ value around, without dissipating it. Hence, these two terms will vanish after taking the ensemble average. And the end result is:
\begin{equation}
    \frac{d}{d t}\left\langle\frac{1}{2} C^{2}\right\rangle=-D\left\langle(\nabla C)^{2}\right\rangle.
\end{equation}
As we see, the dissipation rate of the scalar variance $\langle C^{2}\rangle$ is equal to $\chi=2D\langle(\nabla C)^2\rangle$. And the effective diffusivity of dye can be estimated by:
\begin{equation}
    D = \frac{d\langle C^2\rangle/dt}{2\left\langle(\nabla C)^{2}\right\rangle}.
\end{equation}

In Fig. S3c, we show the effective diffusivity $D$ of the passive dye as a function of bacterial volume fraction $\phi_b$. Error bars indicate the standard deviation of $D$ calculated at different times during the transient mixing. We see the effective diffusivity increases non-linearly with $\phi_b$, suggesting the bacterial activity increases the molecular diffusion of dye. Meanwhile, non-motile bacteria (red star) do not increase effective diffusivity of dye. We do notice that the estimated effective diffusivity of dye is an order of magnitude higher than its actual diffusivity or diffusion coefficient.

\section{Particle tracking velocimetry (PTV)}
We dispersed 100 $\mu$m polyethylene (PE) fluorescent particles on the surface of the fluid. Those particles are buoyant, since then they remain on the top of the surface during the experiments. Particle positions were recorded using CMOS camera (IO Industries, Flare 4M180), operating at 30 fps and 1140 $\times$ 1140 pixel resolution, in a 3.6 cm $\times$ 3.6 cm region. There were on average 2000 particles in the field of view per frame. We record up to 18,000 images, which produces roughly  20,000,000 accurate particle velocities. These velocities were then phase averaged and interpolated on a spatial grid to obtain a velocity map. Each phase of the velocity map was calculated using approximately 150,000 particle velocities, hence highly accurate. 

In Fig. S4, we show the velocity magnitude field and vorticity field data measured using PTV experiments. The data have 475 $\times$ 475 grid points in a region of 30 mm $\times$ 30 mm. There are approximately 16 grid points in each millimeter of the velocity map. This allows us to perform spatial differentiation of the velocity field with errors of only a few percent. 

\section{Finite-time Lyapunov exponent (FTLE)}
We numerically integrated the trajectories of virtual particles in the velocity field (map), using a standard tool \cite{Onu2015}. From the hypothetical trajectories we obtained the flow map, $\mathbf{x}=\mathbf{\chi}(\mathbf{x}_0,t)$, which maps a particle's initial position $\mathbf{x}_0$ to its current position $\mathbf{x}$. We then differentiated the flow map $\mathbf{\chi}(\mathbf{x}_0,t)$ with respect to its initial configuration $\mathbf{x}_0$ to get the deformation gradient tensor, $\mathbf{F}_{t_0}^t(\mathbf{x})=\partial\mathbf{\chi}/\partial\mathbf{x}_0$, and obtained the right Cauchy-Green tensor, via $\mathbf{C}_{t_0}^t(\mathbf{x})=[\mathbf{F}_{t_0}^t(\mathbf{x})]^T\mathbf{F}_{t_0}^t(\mathbf{x})$. Next, we calculated the square root the largest eigenvalue of $\mathbf{C}_{t_0}^t(\mathbf{x})$, denoting it as $\lambda_{t_0}^t(\mathbf{x})$. Finally, the FTLE field can be computed as: $\Lambda_{t_0}^t(\mathbf{x})=\log(\lambda_{t_0}^t(\mathbf{x}))/(t-t_0)$, where $t$ and $t_0$ are in the unit of periods. Different choice of $t_0$ would yield different FTLE results. Here, we chose $t_0$ to be the rising zero of the sinusoidal forcing. There are 2 quantities that can be calculated from this procedure. For $t>t_0$, $\Lambda_{t_0}^t(\mathbf{x})$ was called the forward-time FTLE; for $t<t_0$, $\Lambda_{t_0}^t(\mathbf{x})$ was called the backward-time FTLE. 

These two quantities, tend to be large near the Lagrangian coherent structures (LCSs) of the flow. The backward-time FTLE is large near the attracting LCSs, or unstable manifolds of the flow, which marks the material surfaces that attract nearby fluid parcel trajectories. The forward-time FTLE is large near the repelling LCSs, or stable manifolds of the flow, which marks the border that repel fluid parcel trajectories towards different regions. More details can be found in \cite{Haller2011}.

In Fig. S5 and Fig. S6, we show the experimentally measured forward-time and backward-time FTLE fields for the buffer, active suspensions of different $\phi_b$, and non-motile bacteria suspension. We see as $\phi_b$ increases, the shapes of the flow structures remain similar. But the amplitude of both the forward and backward-time FTLE fields decrease with increasing $\phi_b$, especially in the regions near the flow stable and unstable manifolds. The non-motile bacteria suspension, on the other hand, has much larger FTLE amplitude compared to active suspension of the same $\phi_b$, for both forward and backward (time) FTLE.

In Fig. S7, we show the statistics of the forward and backward (time) FTLE fields. We see the distributions of backward FTLE (Fig. S7a) and forward FTLE (Fig. S7b) are distinct from each other, suggesting the occurrence of temporal reversal-symmetry breaking. The addition of bacteria attenuates higher values of both forward and backward FTLE. But the FTLE distribution at lower values are nearly unaffected by the bacteria. In In Fig. S6c, we show the spatially averaged FTLE $\langle\Lambda\rangle$ as a function of $\phi_b$. We find that both forward and backward $\langle\Lambda\rangle$ decreases non-linearly with $\phi_b$, indicating bacteria is decreasing both fluid elements repulsion and attraction.

%%%%%%%%%%%%% Figure 4
\begin{figure}\label{Fig.S4}
\centering
\includegraphics[width=6.2in]{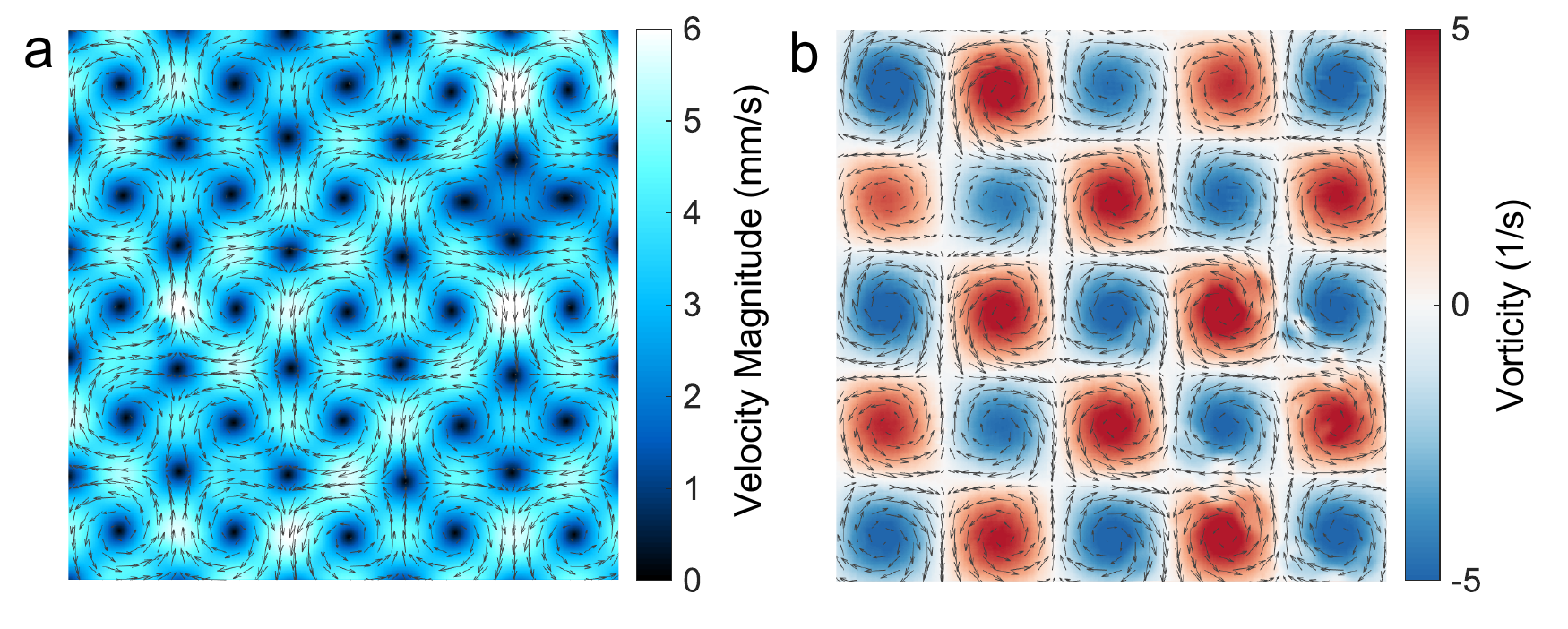}
\caption{Flow field measured from particle tracking velocimetry (PTV), for experiments at Reynolds number $Re=14.0$, path length $p=2.33$. The data is shown in a $5L\times5L$ ($30 \text{mm}\times30 \text{mm}$) region, at its peak magnitude (phase = $\pi/2$). (\textit{a}) Color code indicates the magnitude of the velocity field $|\mathbf{v}|$. Arrows are velocity vector. (\textit{b}) Color code indicates the vorticity field $\mathbf{\omega}$, defined as the curl of velocity field, $\mathbf{\omega}=\nabla\times \mathbf{v}$.}
\end{figure}

%%%%%%%%%%%%% Figure 5
\begin{figure}\label{Fig.S5}
\centering
\includegraphics[width=6.2in]{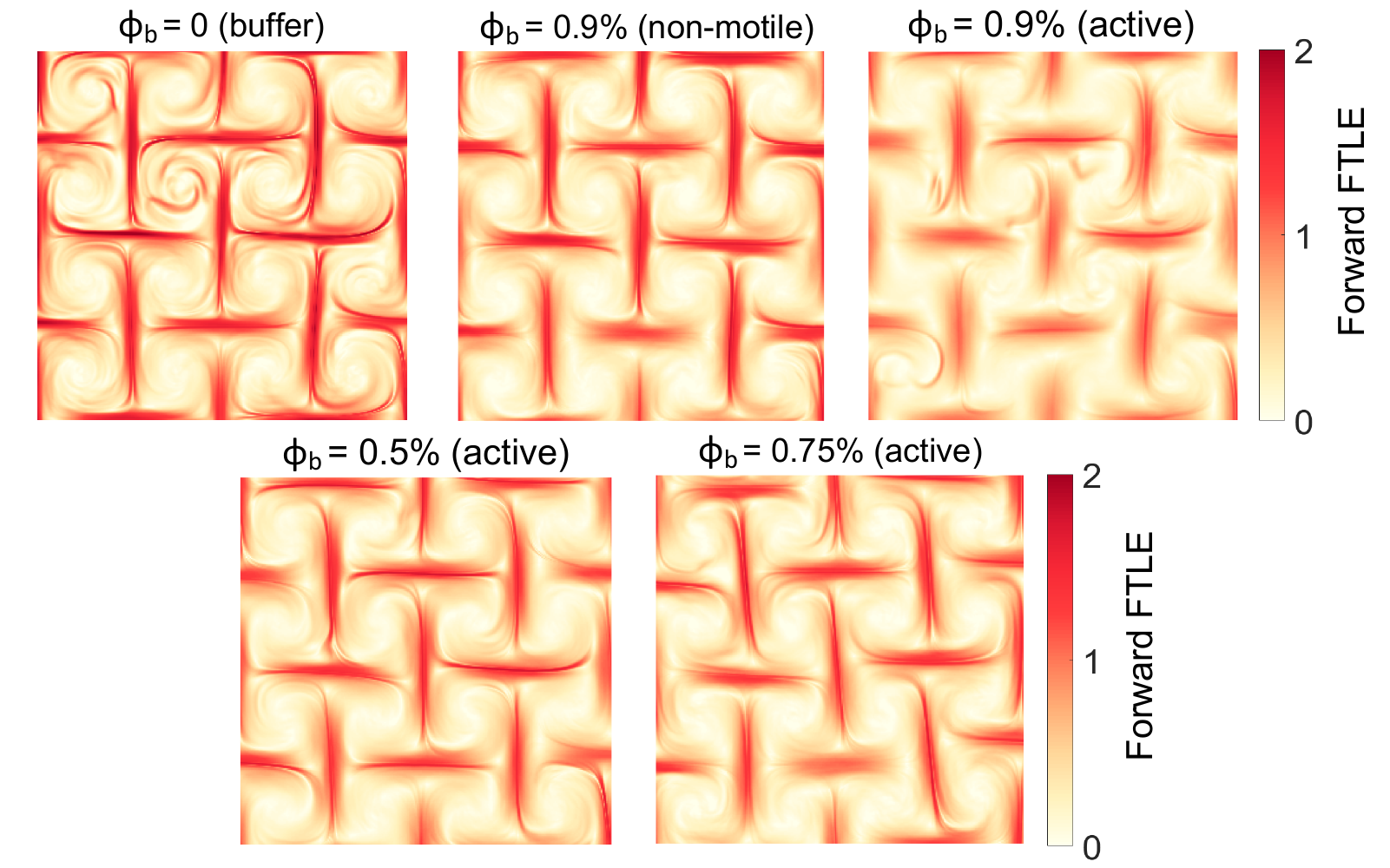}
\caption{The forward (time) finite-time Lyapunov exponent (FTLE) field for the buffer, active suspension of various $\phi_b$, and non-motile bacteria suspension. The ridges of the forward FTLE field marks the repelling Lagrangian coherent structures (LCSs) or the stable manifolds in the flow. Increasing bacterial activity $\phi_b$ systematically attenuates the magnitude of the forward FTLE. The non-motile bacteria suspension has much higher FTLE value compared to active suspension of the same $\phi_b$. The integration time for the calculation of FTLE is $\Delta t$ = 1.5 periods.}
\end{figure}

% Characterization of Lagrangian coherent structures (LCSs) from experimental velocity field measured from PTV. (\textit{a}) 

%%%%%%%%%%%%% Figure 6
\begin{figure}\label{Fig.S6}
\centering
\includegraphics[width=6.2in]{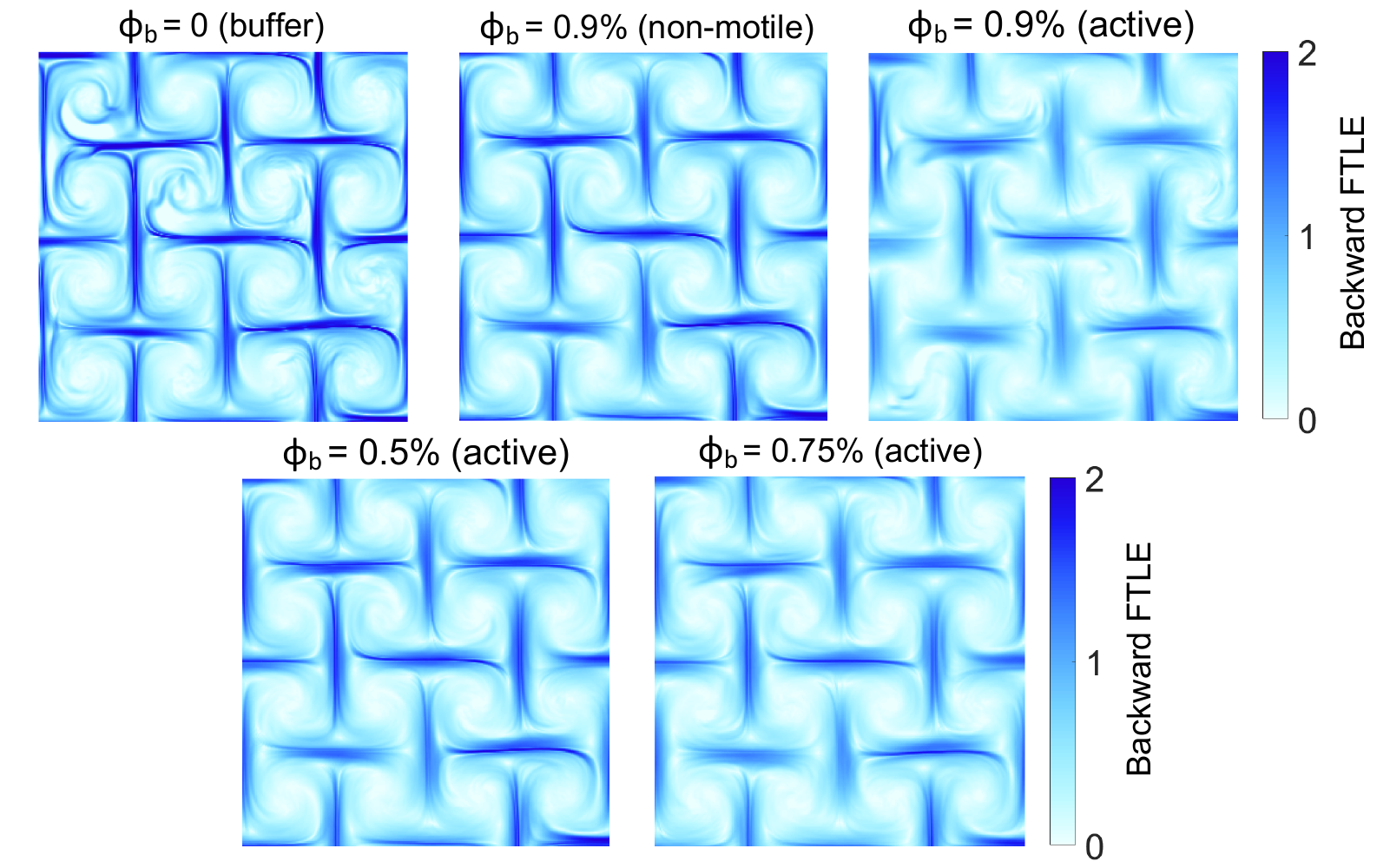}
\caption{The backward (time) finite-time Lyapunov exponent (FTLE) field for the buffer, active suspension of various $\phi_b$, and non-motile bacteria suspension. The ridges of the backward FTLE field marks the attracting Lagrangian coherent structures (LCSs) or the unstable manifolds in the flow. Increasing bacterial activity $\phi_b$ systematically attenuates the magnitude of the backward FTLE. The non-motile bacteria suspension has much higher FTLE value compared to active suspension of the same $\phi_b$. The integration time for the calculation of FTLE is $\Delta t$ = 1.5 periods.}
\end{figure}

%%%%%%%%%%%%% Figure 7
\begin{figure}\label{Fig.S7}
\centering
\includegraphics[width=5.4in]{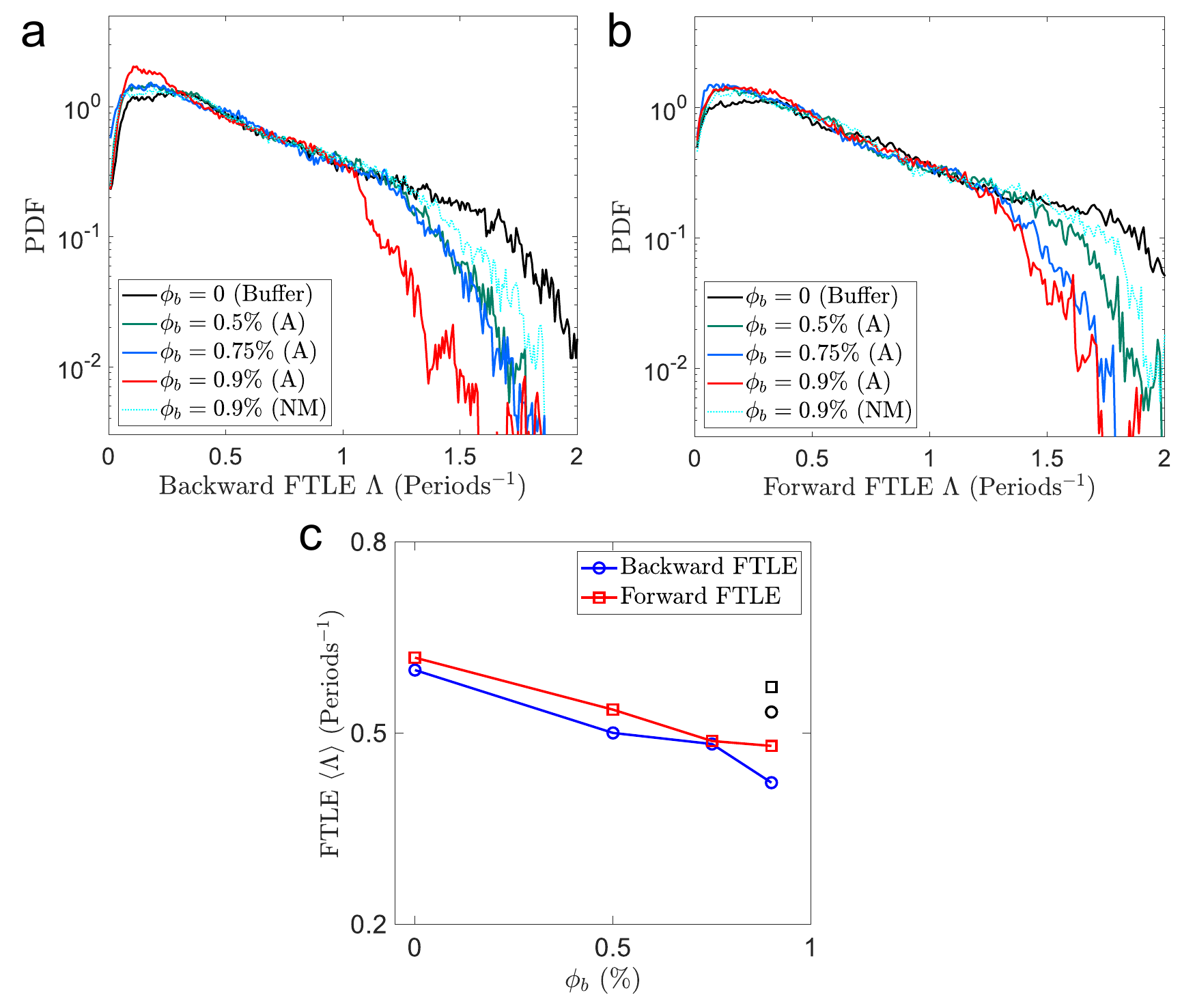}
\caption{The statistics of finite-time Lyapunov exponent (FTLE) fields. (\textit{a}) The probability distribution of backward-time FTLE, characterizing fluid element attraction, for the buffer, active suspensions (A), and non-motile bacteria suspension (NM). (\textit{b}) The probability distribution of forward-time FTLE, characterizing fluid element repulsion. (\textit{c}) The spatial average FTLE as a function of $\phi_b$, for both forward and backward time data. The data suggest increasing $\phi_b$ attenuate both fluid element repulsion and attraction, i.e. fluid element stretching in forward and backward time. The non-motile bacteria suspension (black markers), shows less attenuation in both forward and backward FTLE compared to active suspension of the same $\phi_b$.}
\end{figure}

\section{Spatial power spectrum of the concentration field}
We calculated the two-dimensional (2D) Fast Fourier Transform (FFT) of the dye concentration with uniform rectangular window (i.e. no window), denoting it as $\hat{C}(\mathbf{k})$. The the 2D power spectrum in the polar coordinate is defined as: $E(k,\phi)=|\hat{C}(\mathbf{k})|^2$. The 1D power spectrum of the concentration field was then calculated by: $E_c(k) = \int_{0}^{2\pi} E(k,\phi)k \,d\phi$. Note that $E_c(k)$ is not the circular average of $E(k,\phi)$. Rather, it is the circular average times a prefactor $2\pi k$. By multiplying the prefactor, the integral of $E_c(k)$ over all wavenumber yields the scalar variance $\langle C^2\rangle=\int_0^\infty E_c(k)\,dk$.

In Fig. S8a and b, we show the 2D power spectra $P=E(k,\phi)$ in logarithmic scale, for the buffer ($\phi_b=0$) and active suspension ($\phi_b=0.9\%$). We find the contours for the buffer spread out to much higher wavenumbers, relative to the active suspension. This suggests there are more small-scale or finer structures in the buffer, which have higher wavenumbers. On the other hand, the contours for active suspensions are constraint to lower wavenumbers, corresponding to large-scale or coarser structures. As mixing progresses, big chunks of dye are stretched and folded to produce smaller or finer structures. Since dye structures are larger in the active suspension, it suggests mixing is hindered by the bacteria.

In Fig. S8c and d, we show the 1D power spectra $E_c(k)$ for the buffer ($\phi_b=0$) and all active suspensions ($\phi_b=0.5\%,\ 0.75\%,\ 0.9\%$), measured at $N=100$ and $N=300$, respectively. At earlier time of $N=100$, all spectra collapse onto each other except at small scale for $k>k_\eta$ ($k_\eta$ is the viscous cutoff scale, see Main Text). At later time of $N=300$, we see a consistent increase in the spectral power of $E_c(k)$ with $\phi_b$. Since total spectral power of $E_c(k)$ is the scalar variance, it means that there are more ``unmixedness'' in the active suspension. A comparison of Fig. S8c and d, we find how the effects of bacteria activity propagate from the small to large scales with time, and eventually affect all scales of the flow.

\section{Two-point autocorrelation function of concentration field}
The spatial two-point autocorrelation function $R_c(r_y)$ is defined as $R_c(r_y)=\langle C(y)C(y+r_y)\rangle$, where $\langle\cdot\rangle$ denotes the ensemble average. The autocorrelation function was not computed from its definition in the real space. Rather, it was calculated from the Fourier space through convolution theorem. Recall the 2D Fourier transform of the concentration field, $\hat{C}(\mathbf{k})=\hat{C}(k_x,k_y)$, in the previous section. The average of $\hat{C}(k_x,k_y)$ in the $k_x$ direction gives us $\widetilde{C}(k_y)$. By the convolution theorem, the Fourier transform of the autocorrelation function $\mathcal{F}\{R_c(r_y)\}=\widetilde{C}(k_y)\widetilde{C}^*(k_y)=|\widetilde{C}(k_y)|^2$, where $\mathcal{F}\{\cdot\}$ denotes the Fourier transform, and the superscript $*$ denotes the complex conjugate. The autocorrelation function $R_c(r_y)$ can be finally obtained from the inverse Fourier transform of $|\widetilde{C}(k_y)|^2$.

The spatial-temporal two-point autocorrelation function $R_c(r_y,r_t)$ is defined as $R_c(r_y,r_t)=\langle C(y,t)C(y+r_y,t+r_t)\rangle$. Again, it is calculated via Fourier transform rather than its definition in the real space. We have the time-dependent concentration field $C(\mathbf{x},t)$. We calculated the space-time Fourier transform of $C(\mathbf{x},t)$, denoting as $\hat C(\mathbf{k},\omega)$. Following the same procedure, we have the averaged Fourier transform $\widetilde{C}(k_y,\omega)$. And the spatial-temporal autocorrelation $R_c(r_y,r_t)$ is the inverse 2D Fourier transform of $|\widetilde{C}(k_y,\omega)|^2$. More details about the calculation can be found in \cite{Park2016}.

In Fig. S9, we show the spatial-temporal autocorrelation functions $R_c(r_y,r_t)$ for the buffer ($\phi_b=0$) and all active suspensions ($\phi_b=0.5\%,\ 0.75\%,\ 0.9\%$). As $\phi_b$ increases, the spacing between nearby contours in the $r_t$ direction also increases, indicating the decorrelation time or lifetime of the dye structures goes up with $\phi_b$. We also notice the structures decorrelate at much shorter distance in the active suspensions, particularly for $r_y>2L$. This is because the stochastic process of bacteria activity breaks the spatial periodicity of the dye structures (discussed in the Main Text). 

\section{Differential entropy of concentration gradient PDFs}
We calculated the PDF of the magnitude of concentration gradient, denoting it as $p(G)$. The differential entropy is a continuous extension of the Shannon entropy, the measure of the information of a random variable, which is defined as:
\begin{equation}
    H(p)=\int_{0}^{\infty}-p(G)\log p(G)\,dG
\end{equation}
More details about the differential entropy can be found in \cite{Jaynes1957}.

In Fig. S10, we show the PDFs of the concentration gradient magnitude, at several times, for the buffer ($\phi_b=0$), and active and non-motile bacteria suspensions ($\phi_b=0.9\%$). We find for the buffer and non-motile cases, it takes roughly 100 periods for the PDF to reach an invariant form. However, for active suspension, the time it takes is around 200 periods. The shape of the invariant form for the active suspension is also distinct from the buffer and non-motile cases. This invariant form behavior is captured by the differential entropy of the PDFs (Fig. 4\textit{B, Inset}). 

%%%%%%%%%%%%% Figure 8
\begin{figure}\label{Fig.S8}
\centering
\includegraphics[width=5.4in]{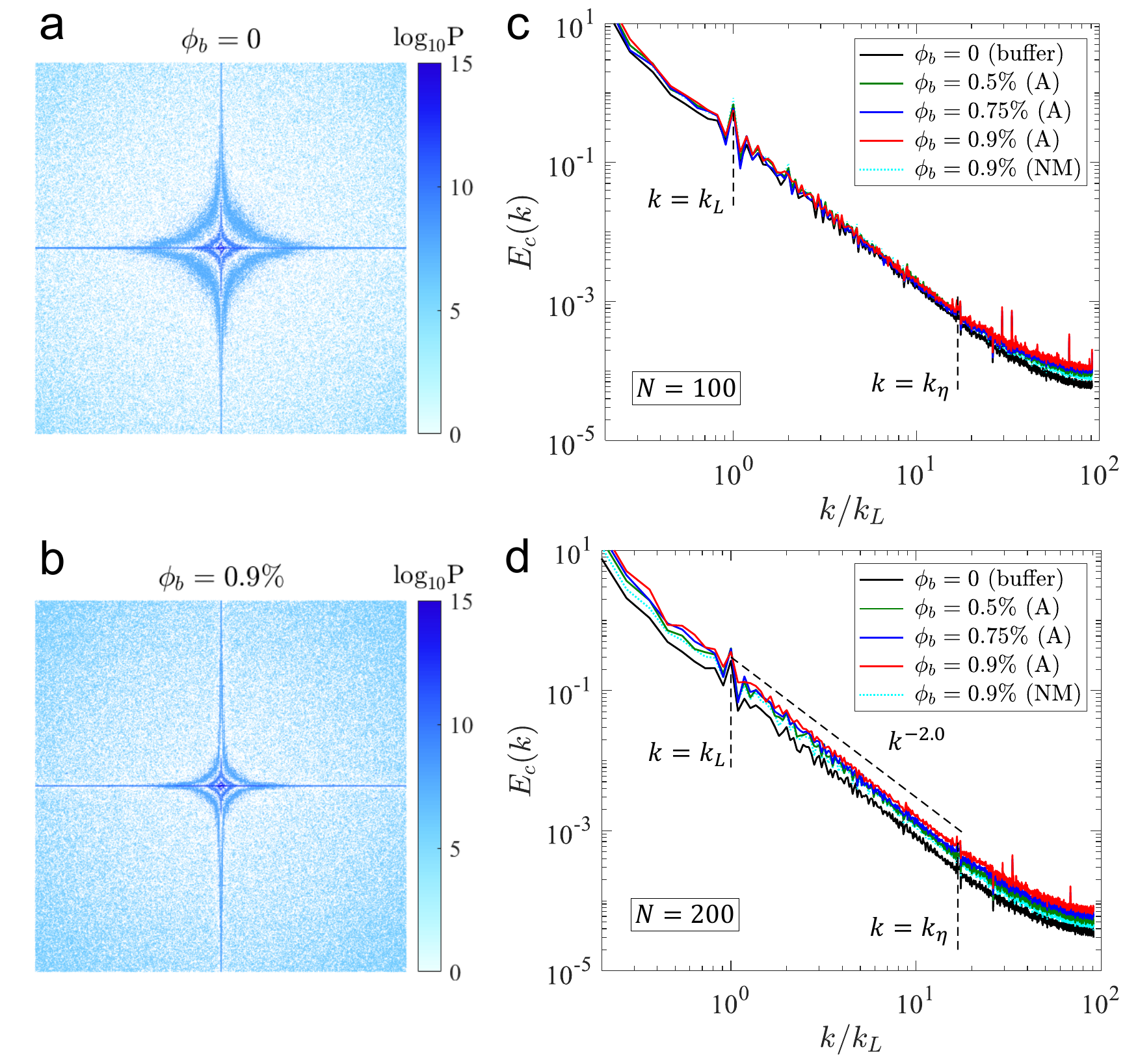}
\caption{Spatial power spectra of the dye concentration field. (\textit{a}, \textit{b}) The two-dimensional power spectra of the concentration field at \textit{N} = 300, for (\textit{a}) $\phi_b=0$ and (\textit{b}) $\phi_b=0.9\%$. The contours of the spectral power for the buffer solution ($\phi_b=0$) spread much further at higher wavenumbers, compared to the active suspension ($\phi_b=0.9\%$), suggesting the structures are much finer (smaller in size or larger in wavenumber) without bacteria. (\textit{c}, \textit{d}) The one-dimensional power spectra $E_c(k)$ measured at (\textit{c}) \textit{N} = 100 and (d) \textit{N} = 200, for the buffer, active suspensions (A), and non-motile bacteria suspension (NM). Two vertical dashed lines mark the energy injection scale $k_L$ and the viscous cutoff scale $k_\eta$. A power law of $E_c(k)\sim k^{-2.0}$ is found all cases, as a characteristic of the flow. At earlier time of \textit{N} = 100, all spectra seem to collapse, except at small scales of $k>k_\eta$. At later time of \textit{N} = 200, the spectra separate from each other at all wavenumbers.}
\end{figure}

%%%%%%%%%%%%% Figure 9
\begin{figure}\label{Fig.S9}
\centering
\includegraphics[width=5.4in]{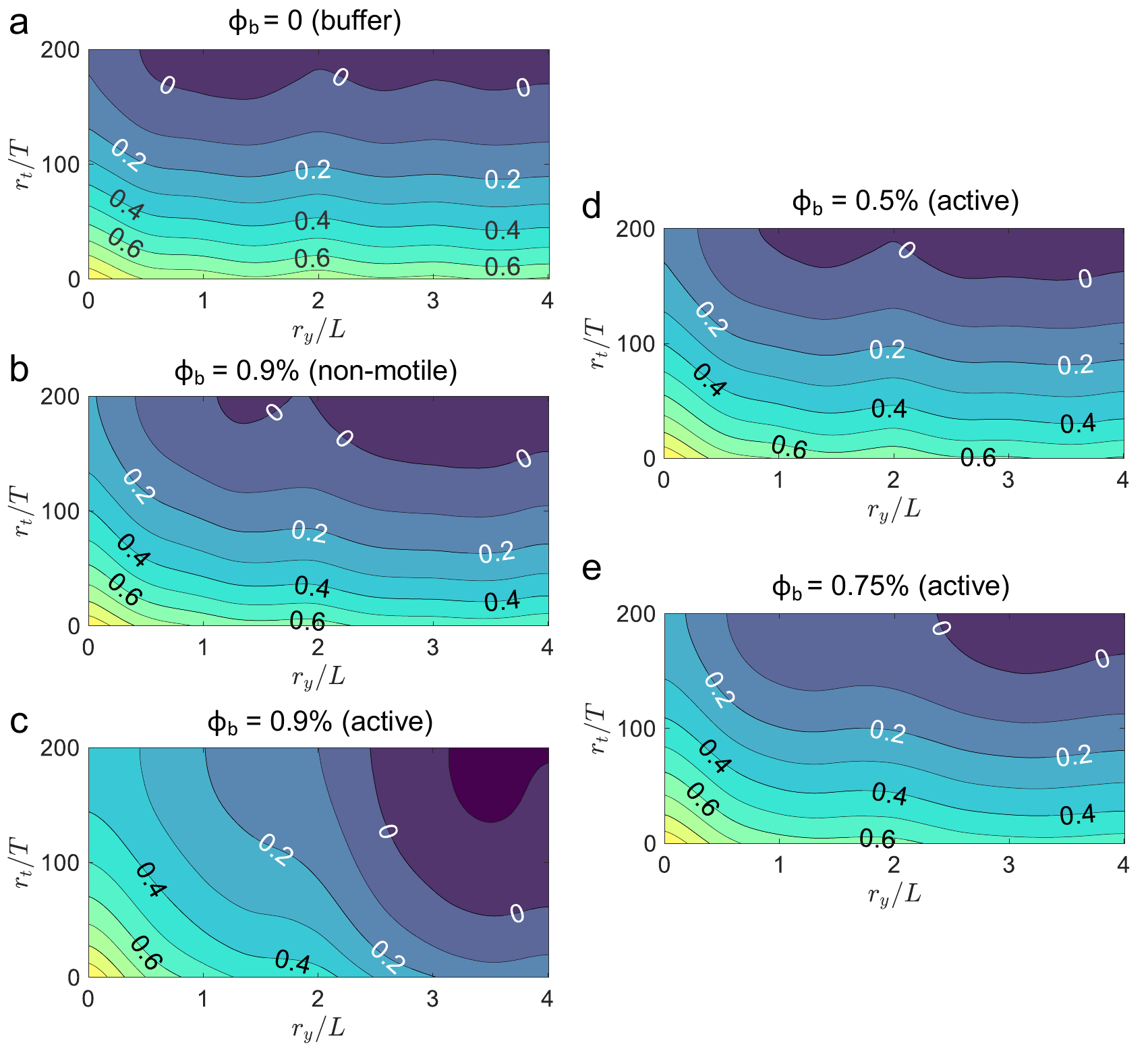}
\caption{The contour plots of spatial-temporal autocorrelation function $R_c(r_y,r_t)$ of the dye concentration field, for (\textit{a}) the buffer of $\phi_b=0$, (\textit{b}) non-motile bacteria suspension of $\phi_b=0.9\%$, (\textit{c}) active suspension of $\phi_b=0.9\%$, (\textit{d}) active suspension of $\phi_b=0.5\%$, and (\textit{e}) active suspension of $\phi_b=0.75\%$. We see an increase in temporal correlation and decrease in long-range spatial correlation with increasing $\phi_b$. Non-motile bacteria suspension does not follow this trend compared to active suspension of the same $\phi_b$.}
\end{figure}

%%%%%%%%%%%%% Figure 10
\begin{figure}\label{Fig.S10}
\centering
\includegraphics[width=5.4in]{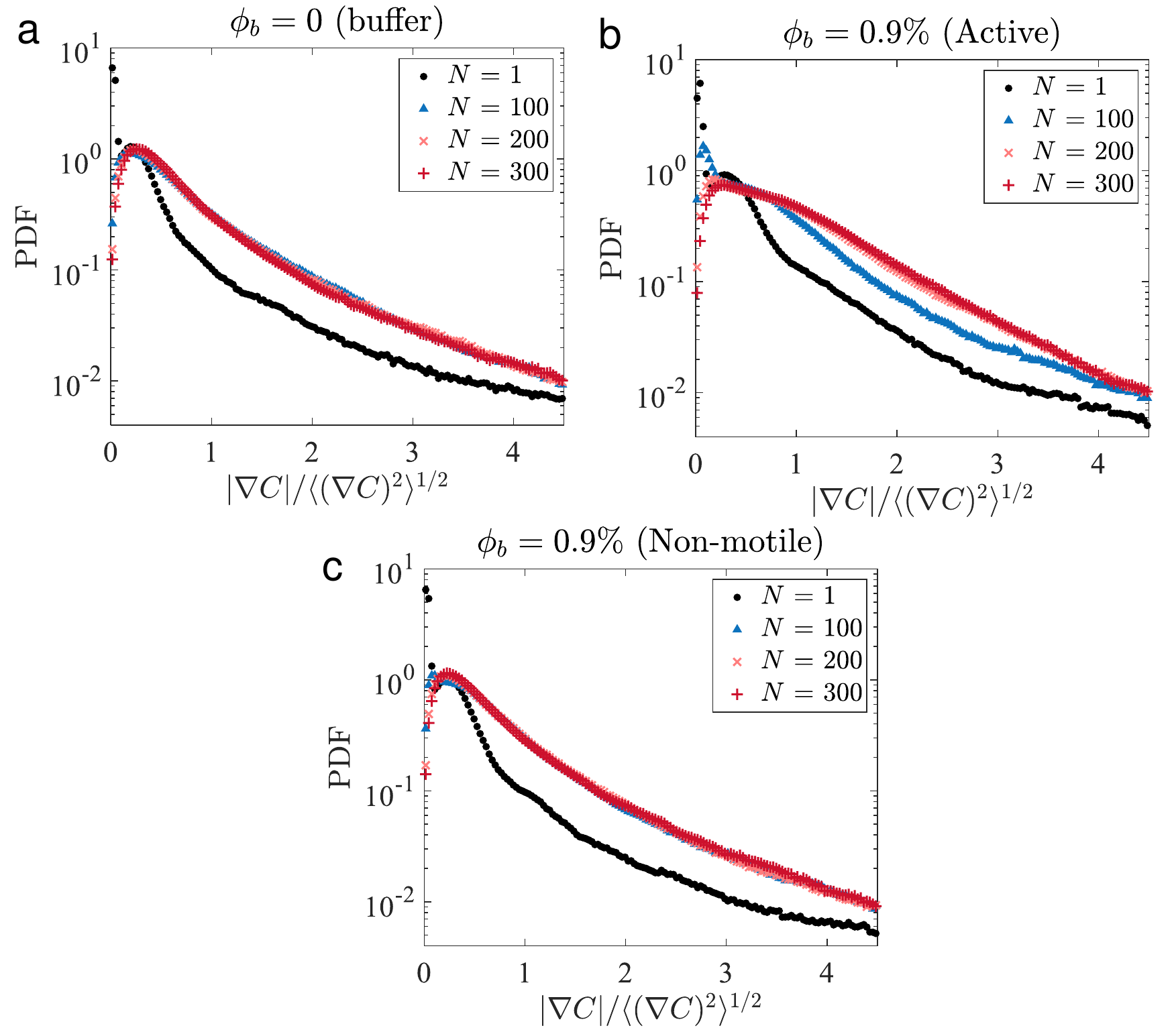}
\caption{Time development of the invariant form (persistent pattern) of the PDF of concentration gradient. (\textit{a}-\textit{c}) The PDF of the magnitude of concentration gradient $|\nabla C|$, normalized by the root means square gradient $\langle(\nabla C)^2\rangle^{1/2}$, at several different times for (\textit{a}) buffer solution ($\phi_b=0$), (\textit{b}) active bacteria suspension ($\phi_b=0.9\%$) and (\textit{c}) non-motile bacteria suspension ($\phi_b=0.9\%$).}
\end{figure}

\section{Measurement of bacteria swimming speed}
To quantify the bacterial activity over our experimental window, we measured the effective diffusivity and swimming speed of \textit{E. coli} in a PDMS microfluidic channel, as shown in Fig. S11a. The channel has a height of 80 $\mu$m and a maximum width of 720 $\mu$m. The swimming motion of \textit{E. coli} were observed under microscope at 0, 10, 20, 30, 60, 120, 240 minutes after they were submerged into the buffer. We tracked the \textit{E. coli}, and obtained the mean square displacement (MSD) of them, as shown in Fig. S11b. The MSD results show that the activity of \textit{E. coli} changes very little in the time period of 4 hours, even without any nutrient sources in the buffer. The time window of our dye mixing is only 30 minutes, during which the bacterial activity should be nearly a constant.

We then fitted the MSD data to a generalized Langevin equation, that is:
\begin{equation}
    \operatorname{MSD}(\Delta t)=4 D_{\text {eff }} \Delta t\left(1-\frac{\tau_0}{\Delta t}\left(1-\mathrm{e}^{-\frac{\Delta t}{\tau_0}}\right)\right),
\end{equation}
where $\tau_0$ is a transition time scale from the ballistic to the diffusive regime. From the fitting of this equation, we obtained the \textit{E. coli} effective diffusivity, $D_{\text {eff}}$, which is shown in Fig. S11c. Again, we see little change in effective diffusivity over time, especially in the first hour of the observation. We show the swimming speed. Next, we obtain the swimming speed of from the effective diffusivity, using the following equation \cite{Howse_PRL_2007}:
\begin{equation}
    D_{\mathrm{eff}}=D_0+ \frac{V_s^{2}}{4D_{R}},
\end{equation}
where $D_0$ and $D_{R}$ are the transnational and rotational diffusivity of \textit{E. coli}, respectively; and $V_s$ is the \textit{E. coli} swimming speed. The swimming speed data is shown in Fig. S11d. We find that the swimming speed is around 10 $\mu$m/s, and is nearly a constant in the first hour of observation.

%%%%%%%%%%%%% Figure 11
\begin{figure}\label{Fig.S11}
\centering
\includegraphics[width=6.2in]{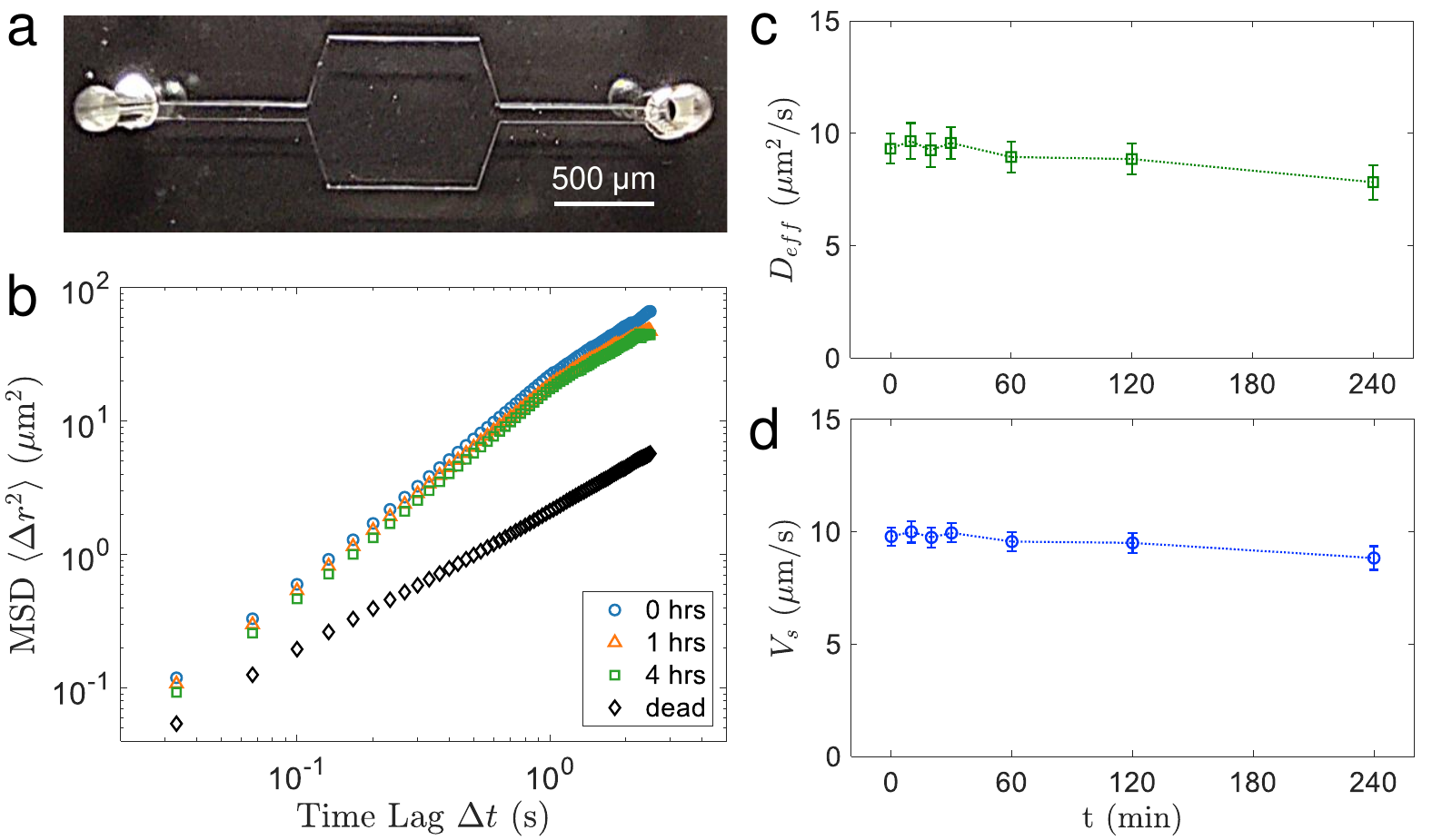}
\caption{Characterization of \textit{E. coli} activity. (\textit{a}) The PDMS microfluidic device that was used to measure the activity of \textit{E. coli}. (\textit{b}) The mean square displacement (MSD) of \textit{E. coli} as a function of time, showing little change in the period of 4 hours. (\textit{c}) The effective diffusivity of \textit{E. coli} as a function of time. (\textit{d}) The swimming speed of \textit{E. coli} calculated from the effective diffusivity as a function of time.}
\end{figure}

%%%%%%%%%%%%% Simulation
\section{Simulation of swimmer transport}
% The flow fields in numerical simulations are defined by a 2D Taylor-Green vortex stream function:
% \begin{equation}\label{eqn.streamfunc}
%     \psi = (UL/\pi)\sin(\pi x/L)\sin(\pi y/L)\sin(2\pi ft)
% \end{equation}
% where $U$ is the maximum velocity, related to the root mean square (RMS) velocity by $\bar{U}=U/\sqrt{8}$; and $L$ is the vortex size, $f$ is the frequency of the flow. The Reynolds number, $Re=\bar{U}L/\nu$, is controlled to be identical to the experiments. The domain of the simulation is a $6L$ by $6L$ region, with periodical boundary conditions imposed on all boundaries.

In Fig. S12a and b, we show the velocity magnitude and vorticity fields of the simulation flow. A comparison of Fig. S12 to the experimentally measured velocity fields in Fig. S4 will reveal that the flow structures are almost the same. In Fig. S12c and d, we show the forward and backward-time FTLE fields of the simulation flow. We find very similar flow structures near the stable and unstable manifolds (repelling and attracting LCSs), when compared to the experimental FTLE fields in Fig. S5 and Fig. S6. These results suggest the stream function we used is a legitimate representation of our experimental velocity fields.

In Fig. S13a and b, we show the particle distribution at $N=50$ colored by the cosine and sine function of their swimming angles, respectively. The particles swimming leftward/rightward appear to be blue/red in Fig. S13a, and green in Fig. S13b. The particles swimming upward/downward appear to be green in Fig. S13a, and red/blue in Fig. S13b. We find swimming particles align to the unstable manifolds of the flow (see Fig. S12c). But there is no preference in the alignment parallel/anti-parallel to the flow velocity. The blue/red particles along the manifolds are almost half and half. This is likely due to the fact that the flow has time symmetry: the particles that are parallel to the flow direction in the first half period become anti-parallel in the second half period, and vice versa.

In Fig. S14a, we show the variance of the number density of the swimming particles. Initially, since all particles are uniformly distributed, the variance is zero. At an earlier time around $N=40$, the variance shows the most intense peak, suggesting strong accumulation of particles along the unstable manifolds. The variance drops to lower values and almost reaches a steady state for $N>120$. The transient behavior of the swimmers is very similar to an under-damped dynamical system. We show that the accumulation of swimmers at earlier times (the first peak) is the strongest.

In Fig. S14b, we show the maximum of the vertically averaged number density of the swimming particles. Similar to the variance result in Fig. S14a, the maximum average density also shows a strong peak at earlier time around $N=35$. The values then decreases and fluctuates, finally reach a dynamical steady state around $N>100$. Unlike the variance, the maximum is estimated without using all the values in the simulation domain. Hence, the data contain more noise. But the behavior of the maximum average number density is very similar to the variance result.  

%%%%%%%%%%%%% Figure 12
\begin{figure}\label{Fig.S12}
\centering
\includegraphics[width=6.2in]{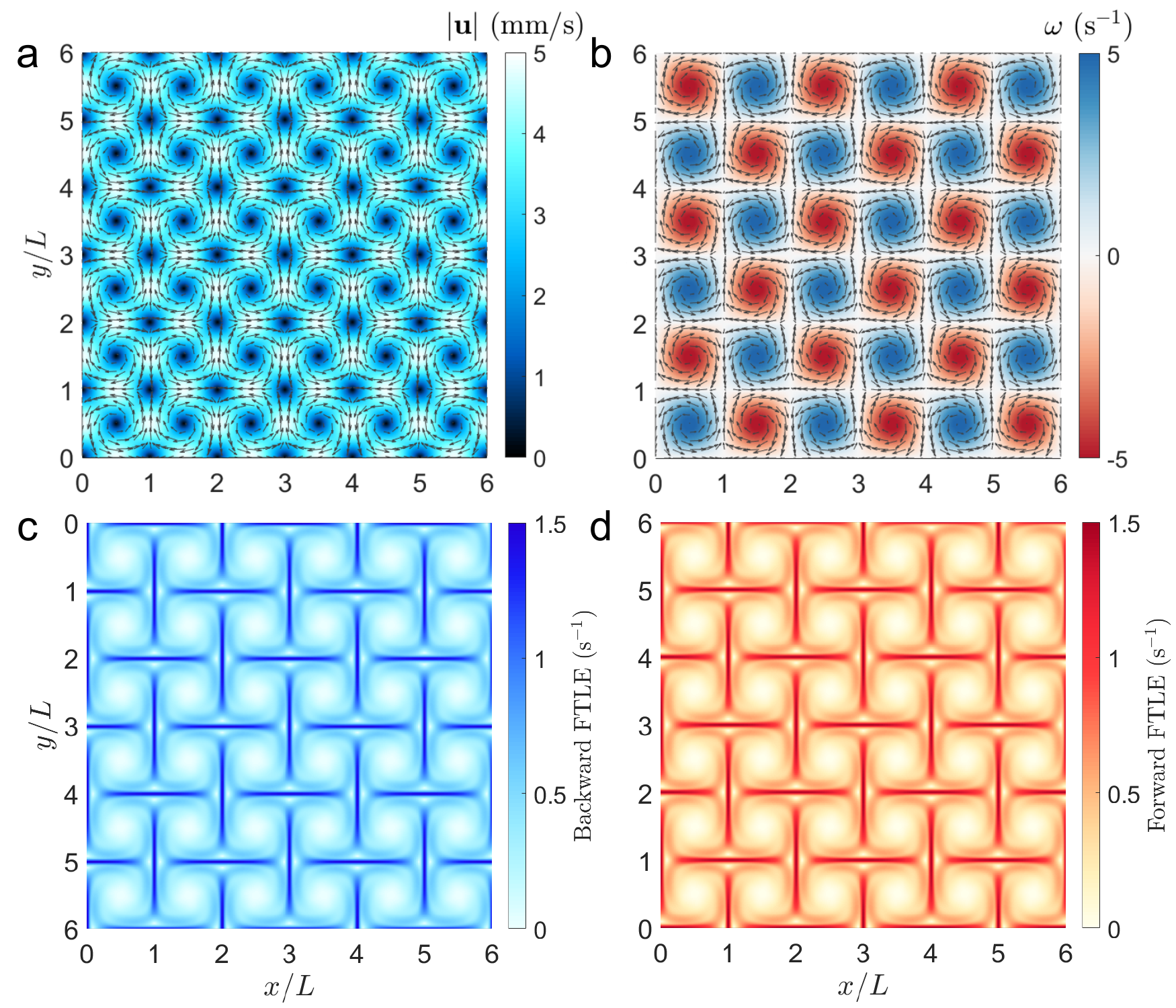}
\caption{Flow field characterization for the stream function used in the numerical simulation, for Reynolds number $Re=14.0$, path length $p=1.94$. (\textit{a}) The magnitude of the velocity field $|\mathbf{v}|$ (colormap), and velocity vector (arrows). (\textit{b}) The vorticity field $\mathbf{\omega}$ (colormap), and velocity vector (arrows). (\textit{c}) The backward-time FTLE field, which marks the finite-time unstable manifolds (attracting LCSs) in the flow. (d) The forward-time FTLE field, which marks the finite-time stable manifolds (repelling LCSs) in the flow. The integration time for both forward and backward time data is $\Delta t$ = 0.5 periods.}
\end{figure}

%%%%%%%%%%%%% Figure 13
\begin{figure}\label{Fig.S13}
\centering
\includegraphics[width=6.2in]{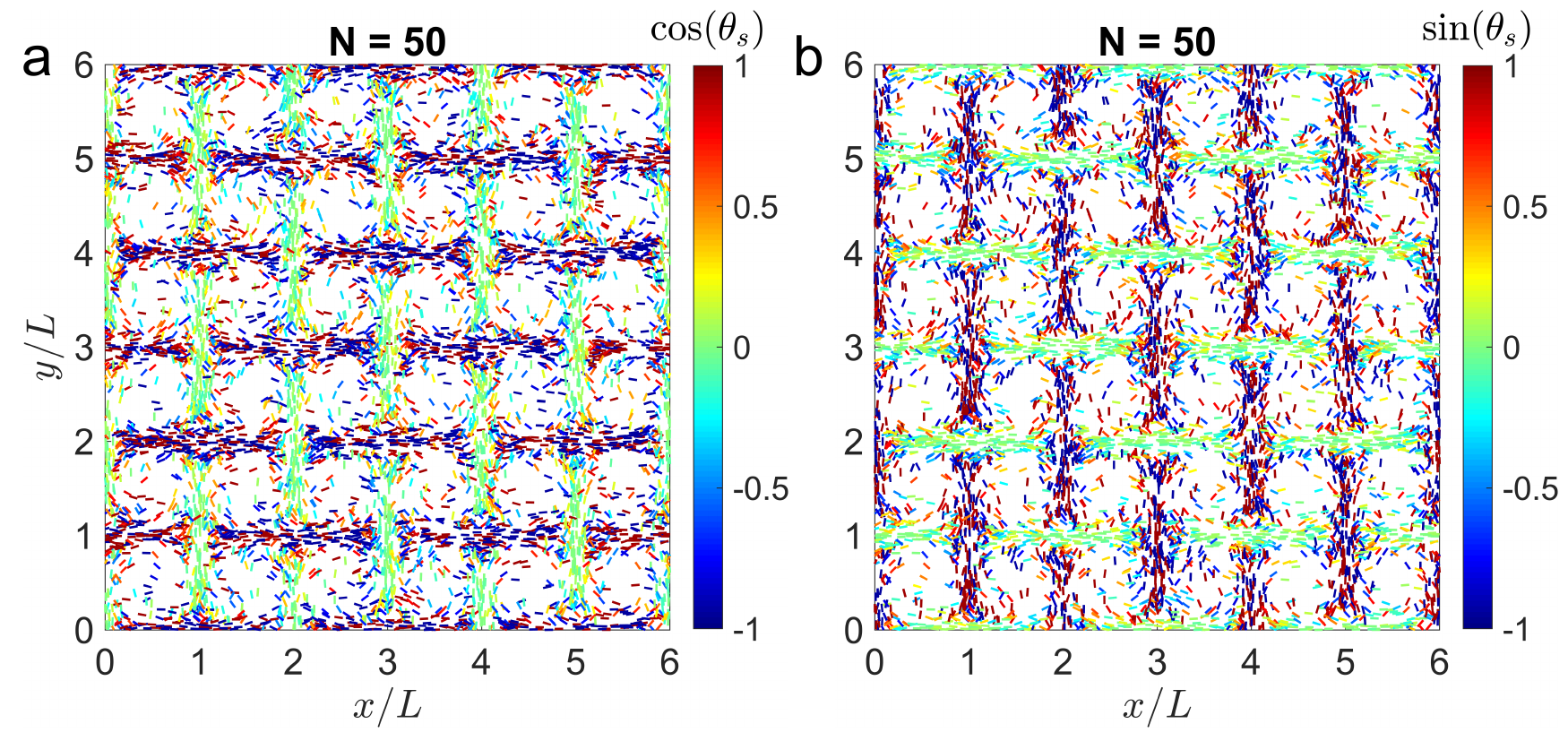}
\caption{Numerical simulation of the pattern formation of elongated swimming particles in the time-periodic flow. (\textit{a}, \textit{b}) Particle distribution in time-periodic flow of Reynolds number $Re=14.0$, path length $p=1.94$, measured at \textit{N} = 50 periods. Swimming particles (visualized as rods) are colored by (\textit{a}) the cosine of swimming angle $\theta_s$ and (\textit{b}) the sine of swimming angle $\theta_s$. Particles that are strongly aligned with the vertical unstable manifolds appear to be green in (\textit{a}), and red/blue in (\textit{b}). Similarly, particles strongly aligned with the horizontal unstable manifolds appear to be red/blue in (\textit{a}), and green in (\textit{b}). The swimmers show no preference in parallel/anti-parallel alignment with the unstable manifolds.}
\end{figure}

%%%%%%%%%%%%% Extensional viscosity
\section{Estimation of local extensional viscosity}
We have argued that the interplay between activity and imposed flow seem to cause bacteria to accumulate along/near the unstable manifolds, or lines of the largest fluid stretching. This is corroborated by our simulations, which show that the accumulation of active swimmers near the hyperbolic regions is about 12 times the initial number density. We speculated that this local accumulation of bacteria can increase the local fluid viscosity, which is a possible underlying mechanism of the hindrance effect in large-scale mixing and transport. However, since the local flow kinematic is extensional in the hyperbolic regions, we believe that a more appropriate fluid property to be used here is the extensional viscosity. The increase in the local extensional viscosity should decrease local fluid stretching, thereby reducing the mixing rate. To better quantify this mechanism, we calculated the increase in extensional viscosity of a suspension of rodlike swimmers, following the analytical results in a previous work on the extensional rheology of active suspensions \cite{Saintillan_PRE_2010}.

The extensional viscosity of the fluid can be tested in an uniaxial extensional flow. The rate-of-strain tensor $\mathbf{D}$ of the flow has the following form:
\begin{equation}
    \mathbf{D}=\dot{\varepsilon}\left[\begin{array}{ccc}
-1 & 0 & 0 \\
0 & -1 & 0 \\
0 & 0 & 2
\end{array}\right],
\end{equation}
where $\dot{\varepsilon}$ is the strain rate. Here, we have $\dot{\varepsilon}>0$ for an extensional flow. It can be shown \cite{Saintillan_PRE_2010} that the total stress of the fluid is
\begin{equation}
    \mathbf{\Sigma} = -p\mathbf{I}+2\mu_0(1+\eta_p\phi_{\text{eff}})\mathbf{D},
\end{equation}
where $p$ is the pressure, $\mu_0$ is the solvent viscosity of the suspension, $\eta_p$ is the intrinsic viscosity of the suspension, and $\phi_{\text{eff}}$ is the effective volume fraction of the rodlike swimmers; here, it is defined as $\phi_{\text{eff}}=\rho_{N} l^{3}$, with $\rho_N$ and $l$ being the number density and length of the swimmers, respectively. The extensional viscosity of the suspension $\mu_{\varepsilon}$ can be defined by the normal stress difference, $\mu_{\varepsilon}=(\Sigma_{11}-\Sigma_{33})/\dot{\varepsilon}$, yielding this Trounton's extensional viscosity in the form of:
\begin{equation}
    \mu_{\varepsilon}=3 \mu_{0}\left(1+\eta_{p}\phi_{\text{eff}}\right).
\end{equation}
Skipping many details of derivation \cite{Saintillan_PRE_2010}, the final expression of the intrinsic viscosity in terms of the P\'eclet number $\mathrm{Pe}$, in the limit of $\mathrm{Pe}\to+\infty$ is:
\begin{equation}
    \eta_{p}=\frac{\pi}{12 \ln (2 \alpha)}\left(\frac{2}{3}+2\left(2+2 \tilde{\sigma}_{0}-\frac{4}{3 \gamma}\right) \mathrm{Pe}^{-1}+\mathrm{O}\left(\mathrm{Pe}^{-2}\right)\right),
\end{equation}
where $\alpha$ is the swimmer aspect ratio, $\gamma=(1-\alpha^2)/(1+\alpha^2)$ is a shape factor, and $\tilde{\sigma}_{0}=\sigma_0/(3k_BT)$ is the normalized force dipole strength of the swimmers, with $k_B$ and $T$ being the Boltzmann constant and Kelvin temperature, respectively. For pushers, $\tilde{\sigma}_{0}<0$; and for puller, $\tilde{\sigma}_{0}>0$. In our experiments, the P\'eclet number is very high. Since then, the expression of the extensional viscosity reduces to:
\begin{equation}\label{ext-vis}
    \mu_{\varepsilon}=3 \mu_{0}\left(1+\frac{\phi \alpha^{2} \pi}{18 \ln (2 \alpha)}\right).
\end{equation}
We can notice the activity stress term $\tilde{\sigma}_{0}$ vanishes in the high P\'eclet number limit, meaning that the particles' contribution to extensional viscosity is very similar to passive particles.

We can now estimate the increase in local extensional viscosity. For particles of the shape of E. coli, the particle aspect ratio is approximately $\alpha\approx4$. According to eqn. \ref{ext-vis}, non-motile bacteria suspension of $\phi_b=0.9\%$  will lead to a uniform increase in extensional viscosity everywhere of about 2.7\%. However, active suspension of the same volume fraction has a 12 times higher local volume fraction of $\phi_b=10.8\%$, leading to an increase of 33.3\% in local extensional viscosity. This explains why the hindrance effect in the non-motile case is not nearly as significant as the active case of the same $\phi_b$.

%%%%%%%%%%%%% Figure 14
\begin{figure}\label{Fig.S14}
\centering
\includegraphics[width=5.4in]{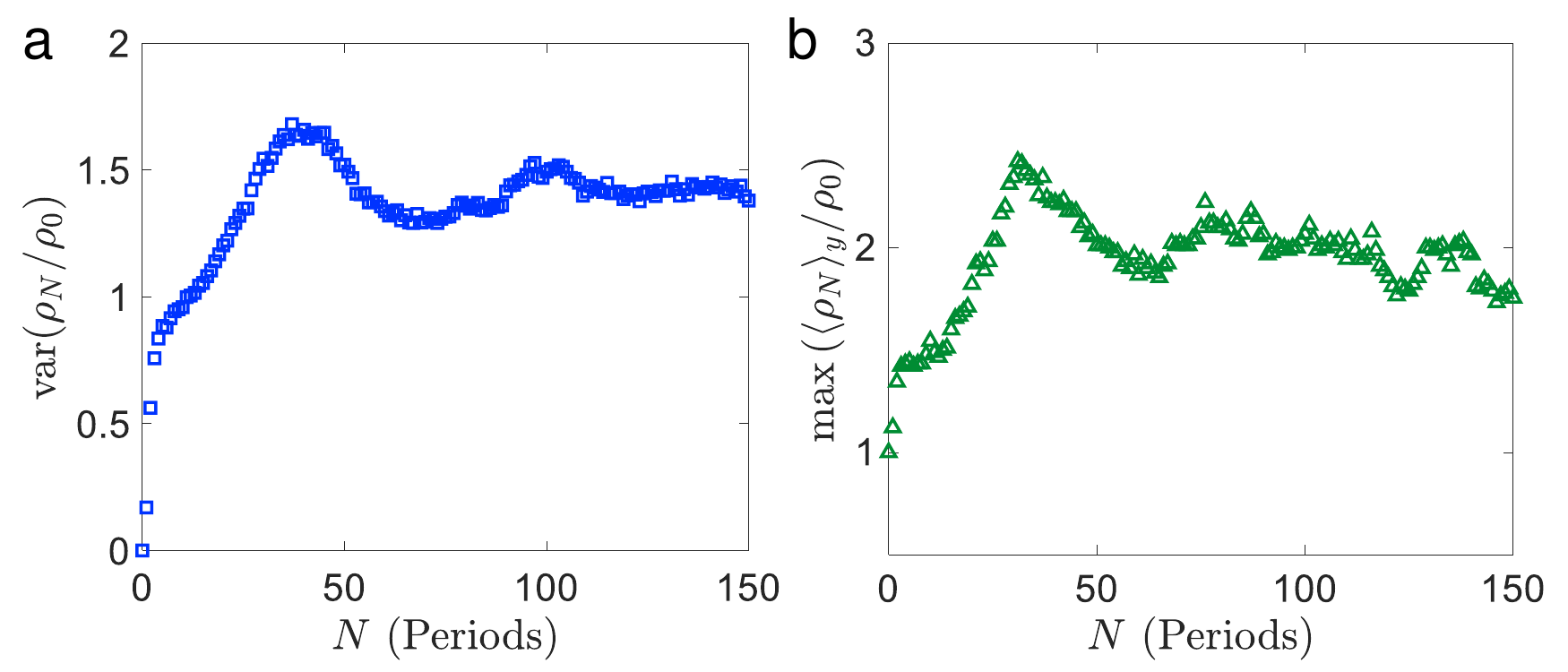}
\caption{Time development of the accumulation of swimming particles. (\textit{a}) The variance of normalized cell number density $\rho_N/\rho_0$ versus time, showing that the strongest accumulation of particles occurs approximately at \textit{N} = 40 periods. (\textit{b}) The maximum of vertically averaged number density $\langle\rho_N/\rho_0\rangle_y$, as a function of time, showing a peak value around \textit{N} = 35 periods. Both data show the transient consists of a strong early accumulation of swimming particles, and a later dispersion of particles.}
\end{figure}

%%% Add this line AFTER all your figures and tables
\FloatBarrier

%%%%%%% Legend for SI Movies
\movie{Stroboscopic video of the dye mixing experiments, for the buffer solution (${\phi_b}$ = 0), active bacteria suspension of different bacterial volume fraction ($\phi_b$ = 0.5\%, 0.75\%, 0.9\%), and non-motile bacteria suspension ($\phi_b$ = 0.9\%). The video shows 400 cycles of the forcing period for totally 2000 seconds. All frames were taken at the rising zero of the sinusoidal forcing (phase = 0). The Reynolds number and path length of all experiments shown here are \textit{Re} = 14.0, \textit{p} = 2.33, respectively.}

\movie{Real time video (3$\times$ faster actual time) of the dye mixing experiments, for the buffer solution (${\phi_b}$ = 0), active suspension of different bacterial volume fraction ($\phi_b$ = 0.5\%, 0.75\%, 0.9\%), and non-motile bacteria suspension ($\phi_b$ = 0.9\%). The video shows 3 cycles of the forcing period for totally 15 seconds, after 300 periods of the forcing (\textit{N} = 301 - 303). The Reynolds number and path length of all experiments shown here are \textit{Re} = 14.0, \textit{p} = 2.33, respectively.}

\movie{Numerical simulations of the transport of ellipsoidal swimming particles. Swimming particles are visualized as rods that are colored by the normalized number density $\rho_N/\rho_0$. All particles are initially uniformly distributed in the whole domain with random orientations. We observe a strong earlier accumulation of particles near the unstable manifolds of the flow around \textit{N} = 40 periods; the number density $\rho_N$ is 10 times larger than the initial density $\rho_0$. The particle distribution becomes more disperse at later times of \textit{N} > 100 periods, as shown by a wider number density pattern.}

%%%%%%%%% reference
\bibliography{supplement}